\documentclass[useAMS,usenatbib]{mn2e}  
 
\usepackage{amsmath}
\usepackage{rotating}
\usepackage{color}
\usepackage{times}
\usepackage{slashbox}
\usepackage{graphicx, subfigure}
\usepackage{multirow}
\usepackage{array}
\usepackage{epsfig}
\usepackage{epstopdf}
\usepackage{float}
\usepackage{gensymb}

\newif\ifpdf
\ifx\pdfoutput\undefined
   \pdffalse
\else
   \pdfoutput=1
   \pdftrue
\fi
\ifpdf
   \usepackage{graphicx}
   \usepackage{epstopdf}
\else
   \usepackage{graphicx}
\fi
 
\newcommand{\kpc}{{\rm\,kpc}}
\newcommand{\kms}{{\rm\,km\,s^{-1} }}
\newcommand{\msun}{{\rm\,M_\odot }}

\title[Stability of Satellite Planes I]{On the Stability of Satellite Planes I: Effects of Mass, Velocity, Halo Shape and Alignment}
\author[Fernando et. al.]{Nuwanthika Fernando$^1$\thanks{E-mail:
fernando@physics.usyd.edu.au (AVR);}, Veronica Arias$^{1,2}$, Magda Guglielmo$^1$, Geraint F. Lewis$^1$,\and Rodrigo A. Ibata$^3$, Chris Power$^4$\\   
\vspace{0.1cm}\\
$^{1}$Sydney Institute for Astronomy, School of Physics A28, The University of Sydney, NSW 2006, Australia\\
$^{2}$Departamento de Fis\'ica, Universidad de los Andes, Cra. 1 No. 18A-10, Edificio Ip, Bogot\'a, Colombia\\
$^{3}$Observatoire de Strasbourg, 11, rue de l'Universit\'e, F-67000, Strasbourg, France\\
$^{4}$ICRAR, University of Western Australia, 35 Stirling Highway, Crawley, Western Australia 6009, Australia
\\}  
\begin{document}
 
\date{Accepted 2016 October 15. Received 2016 October 12; in original form 2016 August 2}
 
\pagerange{\pageref{firstpage}--\pageref{lastpage}} \pubyear{2016}
 
\maketitle
 
\label{firstpage}
 
\begin{abstract}
The recently discovered vast thin plane  of dwarf satellites  orbiting the Andromeda Galaxy (M31) adds
to the mystery of the small scale distribution of the Local Group's
galaxy population. Such well defined planar structures are apparently rare
occurrences in cold dark matter cosmological simulations, and  
we lack a coherent explanation of their formation and existence.
In this paper, we explore the long-term survivability of thin planes of dwarfs  
in galactic halos, focusing, in particular, on systems mimicking the  
observed Andromeda distribution.
The key results show that, in general, planes of dwarf galaxies are  
fragile, sensitive to the shape of the dark matter halo and other  
perturbing effects. In fact, long lived planes of satellites only exist  
in polar orbits in spherical dark matter halos, presenting a challenge to  
the observed Andromeda plane which is significantly tilted with respect
to the optical disk. Our conclusion is that, in standard cosmological models,  
planes of satellites are generally short lived, and hence we must be located at  
a relatively special time in the evolution of the Andromeda Plane, lucky enough to  
see its coherent pattern. %This conclusion, while in itself unsatisfactory, is made more so with the presence of the polar plane in our own Milky Way.  
\end{abstract}
 

\newcommand{\veronica}{\color{red}}
\newcommand{\whoblue}{\color{blue}}

\section{Introduction}\label{Intro}
For decades, the distribution of the Local Group's satellite galaxy
population has presented a puzzle. This began with the discovery of the
alignment of a number of Milky Way (MW) satellite galaxies with the
orbital plane of the Magellanic Clouds \citep{Lynden-Bell1976},
identifying a vast polar structure in the Milky Way, with a second similar grouping
noted by \citet{Lynden-Bell1982}, containing Fornax, Leo I, Leo II and
Sculptor.
More recently,
\citet{Kroupa2005} showed that the 11 brightest satellites of the MW
are on a plane with a thickness of 20 \kpc\ \citep{Pawlowski2012b,Pawlowski2014}, and aligned with the pole of the Galaxy. The
recent analysis of the Pan-Andromeda Archaeological Survey (PAndAS)
 resulted in the accurate distance measurements to the known population of M31
satellites using a sensitive homogeneous method \citep{Conn2012}, revealing  
their three-dimensional distribution.
As shown in \citet{Ibata2013} about 15 of the 30 known satellites that orbit M31 form an extremely thin plane, with a thickness of 12.6$\pm$0.6 \kpc, but with an overall extent of $\sim$200 \kpc. The significance of the presence of this plane was boosted with the discovery of kinematically coherent orbital motion for the plane members, with the southern most satellites moving towards us, while those in the north are moving away (correcting for the motion of M31). The possibility that this plane represents a chance alignment appears to be highly improbable, with a resulting significance of the presence of this planar structure of approximately 99.998 \%\ \citep{Conn2013,Ibata2013}.  
 
From our vantage point, the Andromeda plane is seen edge-on, and is approximately aligned with the pole of our own Galaxy, being significantly tilted with respect to Andromeda's optical disk \citep{Ibata2013,Shaya2013}.  
The existence of additional planes in the Local Group was noted by \citet{Tully2015}, who
finds a second plane in the M31 system. They also find two parallel planes in the Centaurus A group. % *******  
 
There are various open questions in our understanding of planes of satellite galaxies, in particular,  
how such a structure forms and how stable it is over a cosmologically significant period of time; these are the questions we address in this paper.  
We begin in Section~\ref{BGround} by giving an
overview of the recent discussions and
debates the issues of satellite planes, while   
Section~\ref{NumMod} introduces the numerical model used to describe
both the MW and M31 potentials.  
In Sections~\ref{MW_pl} and~\ref{Stb_VelMass}, we describe the stability tests for orbiting satellite
galaxies in a plane, and examine how the MW and satellite mass and velocity
 can influence a plane of satellite.
 In Section~\ref{Stb_NSH}, we study the influence of the properties of the host galaxy's dark halo on
a planar structure, focusing in particular upon the orientation of plane structure with regards to flattening of the potential. Section~\ref{Discn} presents our comparison of simulations with observations and conclusions on the longevity of planes of satellites.
 
﻿\section{Background}\label{BGround}

In the decades since the discovery of MW's generally asymmetric distribution of satellite galaxies, a significant effort has been put into understanding its existence in light of the standard $\Lambda$ Cold Dark Matter ($\Lambda$CDM) cosmology. A search for coherent orbits within numerical simulations of structure formation reveals an overall isotropic distribution for satellite galaxies, with plane-like structures occurring in about 2\% of MW/M31-like halos \citep{Boylan-Kolchin2009, Bahl2014}. The chance of such occurrences become even lower when looking for structures with properties similar to the M31 system \citep{Ibata2014}, suggesting that strongly anisotropic distribution of satellites are not a natural phenomenon within simulations of $\Lambda$CDM cosmology. However, the discrepancies between dark matter simulations and the observations in the Local Group might be resolved by introducing baryonic physics. `Zoom in' simulations of the Local Group with more detailed hydrodynamical simulations, including supernova feedback, show anisotropic distributions of satellite galaxies around Milky Way-like halos, not dissimilar to those around the MW \citep{Sawala2014}. These results seem to be in agreement with those presented in \citet{Ibata2014b}, where the authors investigate the incidence of planar structures in a larger galaxy population using the Sloan Digital Sky Survey (SDSS). Their analysis suggests that corotating dwarf galaxies might be common, although this result has been challenged by recent papers \citep[e.g.][]{Phillips2015}.  
 
%The potential ubiquity of planes has become murkier.
Repeating the analysis presented in \citet{Ibata2014b}, \citet{Cautun2015a} conclude that while co-rotating satellite pairs are seen, they do not imply the presence of co-rotating planes of satellites. In the same paper, the authors compare the occurrence of co-rotating galaxy pairs in the SDSS to those in the Millennium II cosmological simulations, finding a general agreement between the two. \citet{Cautun2015b} explore the Millennium II (MII) and COCO \citep{Hellwing21042016} dark matter only simulations, extending the search for planes of satellites. Around 10\% of MW/M31-like hosts were found to contain satellite planes, albeit not always as rich in number and as thin as  those observed in M31 (and MW). This suggests that satellite planes  are predicted in $\Lambda$CDM, but with properties  (e.g. -varying thickness, fraction of total satellite population on the plane and radial distribution) that can be different to those observed in the M31/MW. Due to this variety, a search matching only the two prominent planes seen in our Local Group in the simulations may yield results that classify the planes as an extreme rarity.
 
While corroborating findings on the diversity of planes in simulations, \citet{Buck2015} explore how the dark halo environment may help or hinder the presence of planes. Their zoom-in dark matter simulations show  that high-concentration halos are more likely to host planes of satellites, having accreted their satellites and mass at early times. However, the general conclusion is that planes of satellites are a transitory phenomenon. Similar conclusions are drawn by the \citet{Gillet2015}  analysis of CLUES (Constrained Local UniversE Simulations). The planes they find contained 11 satellite galaxies at most, indicating the likely possibility of M31's plane of 15 satellites being an extreme occurrence. About a third of satellites have high velocities that are perpendicular to the planar formation, making their presence in the plane purely coincidental. \citet{Gillet2015} conclude that the observed satellite planes appear to be in agreement with the $\Lambda$CDM but as a transitory feature, created with a few planar satellites and several timely interlopers. Such coincidental placement of galaxies may reduce the number of observed kinematically coherent planes. If this holds true, we are living in a ‘special’ time when these structures can be observed both in M31 and MW.  
 
The question of the formation and evolution of planar populations of dwarfs also remains unanswered. There have been several suggested mechanisms, with the feeding of dwarf galaxies through dark matter filaments taking a prominent position \citep{Libeskind2005, Lovell2011}. \citet{Buck2015} claim that M31-like planes emerged in very early forming host halos, where the satellites are accreted around $z \geq 3$ via filaments that were thinner in early times (see \citet{Pawlowski2012b, Pawlowski2014} for different interpretations). More radical solutions have been proposed as well, e.g: \citet{Kroupa2005} suggest that the MW plane, having origins in tidal dwarf galaxies, formed as a result of larger galaxy interactions \citep{Hammer2013}. However, \citet{Collins2015} find that satellites on and off the plane exhibit no segregating features other than their spatial alignment. This would contradict a merger with a similar mass galaxy as a formation scenario for M31; distinct histories would result in different properties for satellites on and  off the M31 plane. Nevertheless, a recent accretion event of such magnitude is found to be improbable \citep{Angus2011}. \citet{Smith2016} explore the behavior of satellite planes after similar mass mergers and find that about 65- 70 \% survive to the simulation's end, while the rest are destroyed or merged. Satellites positioned in/close to dark matter filaments have a slightly higher survival rate and time through their simulations. Although nothing notable is affected by the variations in the applied cosmology, prograde satellites were found to be more likely to stay as satellites, as opposed to retrograde orbits.  
 
The longevity of a plane of dwarfs in a galactic halo has been found to depend on the orientation of its orbit. \citet{Bowden2013} argue that a thin disc of satellite galaxies can persist over cosmological times if and only if it lies in a plane that is aligned with one of the semi-major or semi-minor axis of a triaxial halo, or in the equatorial or polar planes of a spheroidal halo. In this latter case, perturbations from the disk can act to disperse a satellite plane. In any other orientation, the disc thickness would double on 10 Gyr timescales and so, to get planes as thin as the one observed in M31, these must have been born with what they address as an implausibly small perpendicular scale height.  
 
Despite the intricacies of formation processes of the plane of satellites, their longevity will depend on environmental properties (such as host halo shape and the particular orbits the satellite lie on). In this paper, we explore the effect that halo shape and initial properties of the satellites will have on the stability of planar structures.
    
\section{Numerical Model}\label{NumMod}
 
Our numerical model is based upon two static potentials to represent the M31-MW system. Even though the  
focus of this paper is considering the orbits of dwarfs in the vicinity of M31, we chose not to neglect the perturbing influence of an external large galaxy like the MW.  The potential of M31 consists of three components representing the
halo, disk and bulge in separate equations, as first proposed in
\citet{2006MNRAS.366..996G} and used in \citet{Arias2016}. The dark
matter halo is described as a Navarro, Frenk and White (NFW) potential
\citep{1997ApJ...490..493N} given by
\begin{equation}
\label{EqNFW}
\Phi_{\rm{halo}}(r)=-4{\pi}G\delta_{\rm{c}}\rho_{\rm{c}}{r_{\rm{h}}}^2\left(\frac{r_{\rm{h}}}{r}\right)\ln \left(\frac{r+r_{\rm{h}}}{r_{\rm{h}}}\right)
\end{equation}    
where $r_{\rm{h}}$ is the scale radius, the present day critical
density is $\rho_{\rm{c}}=277.7\,h^2\msun\,{\rm{kpc^{-3}}}$, $h=0.71$ in
the unit of 100 $h\ \ \kms\rm{Mpc^{-1}}$ \citep{2006MNRAS.366..996G}  and
$\delta_{\rm{c}}$ is a dimensionless density parameter. %{\whoblue WHY DENSITY IN M / L2 ?}
The disk component of the potential is given by a Miyamoto-Nagai potential \citep{1975PASJ...27..533M}
\begin{equation}
\label{EqMiyaN}
\Phi_{\rm{disk}}(R,z)=-\frac{\rm{GM_{disk}}}{\left(R^2+\left(r_{\rm{disk}}+{\sqrt{(z^2+b^2)}}\right)^2\right)^{1/2}}
%\Phi_{\rm{disk}}(r)=-2{\pi}G\Sigma_{0}{r^2_{\rm{disk}}}\left[\frac{1-\exp^{-r/r_{\rm{disk}}}}{r}\right]
\end{equation}   
 and the bulge component follows a \citet{1990ApJ...356..359H} profile.
\begin{equation}
\label{EqHernquist}
\Phi_{\rm{bulge}}(r)=-\frac{\rm{GM_{{bulge}}}}{r_{\rm{bulge}}+r}\mbox{.}
\end{equation}
The Milky Way potential contains similar components, a NFW halo
(Eq.~\ref{EqNFW}), a Hernquist bulge (Eq.~\ref{EqHernquist}) and
stellar disk defined as a Miyamoto-Nagai potential (Eq.~\ref{EqMiyaN}). All the parameters we use are listed in Table \ref{tabPot} and are
similar to values used in \citet{Arias2016}.

%\begin{table}
%\begin{tabular}{l l l }
%$\Phi_{\rm{halo}}(r)$ & = & \parbox{58mm}{\begin{flushright}\begin{equation}
%\label{EqNFW}
% -4{\pi}G\delta_{\rm{c}}\rho_{\rm{c}}{r_{\rm{h}}}^2\left(\frac{r_{\rm{h}}}{r}\right)\log \left(\frac{r+r_{\rm{h}}}{r_{\rm{h}}}%\right)
%\end{equation}\end{flushright} }
%\\
%$\Phi_{\rm{disk}}(R,z)$ & = &\parbox{58mm}{\begin{flushright} \begin{equation}
%\label{EqMiyaN}
% -\frac{\rm{GM_{disk}}}{\left(R^2+\left(r_{\rm{disk}}+{\sqrt{(z^2+b^2)}}\right)^2\right)^{1/2}}
%%\Phi_{\rm{disk}}(r)=-2{\pi}G\Sigma_{0}{r^2_{\rm{disk}}}\left[\frac{1-\exp^{-r/r_{\rm{disk}}}}{r}\right]
%\end{equation} \end{flushright}}   
%\\
%$\Phi_{\rm{bulge}}(r)$ & = & \parbox{58mm}{\begin{flushright} \begin{equation}
%\label{EqHernquist}
% -\frac{\rm{GM_{{bulge}}}}{r_{\rm{bulge}}+r}%\mbox{.}
%\end{equation} \end{flushright}}
%\\
%\end{tabular}
%\end{table}

\begin{table*}
\centering
\begin{tabular}{l l p{55mm} p{15mm} l l} %{l l l l l l}
\hline
%\multicolumn{3}{c}{\textbf{M31}}{\textbf{MW}}\\
    &    &    &    & \textbf{M31}              & \textbf{MW}\\
\hline
\hline
& & & & &\\
$\Phi_{\rm{halo}}(r)$ & = & \parbox{58mm}{\begin{flushright}\begin{equation*}
\label{EqNFW}
 -4{\pi}G\delta_{\rm{c}}\rho_{\rm{c}}{r_{\rm{h}}}^2\left(\frac{r_{\rm{h}}}{r}\right)\log \left(\frac{r+r_{\rm{h}}}{r_{\rm{h}}}\right)
\end{equation*}\end{flushright} } & $\rm{r}_{h}$        & $13.5\,\rm{kpc}$    & $24.54\,\rm{kpc}$ \\
& & & $\rm{M}_{halo}$ &  $1.037\times{10^{12}}\,\rm{M}_{\odot}$ & $0.9136\times{10^{12}}\,\rm{M}_{\odot}$ \\
$\Phi_{\rm{bulge}}(r)$ & = & \parbox{58mm}{\begin{flushright} \begin{equation*}
\label{EqHernquist}
 -\frac{\rm{GM_{{bulge}}}}{r_{\rm{bulge}}+r}%\mbox{.}
\end{equation*} \end{flushright}} & $\rm{M}_{bulge}$    & $2.86\times{10^{10}}\,\rm{M}_{\odot}$   & $3.4\times{10^{10}}\,\rm{M}_{\odot}$ \\
& & & $\rm{r}_{bulge}$    & $0.61\,\rm{kpc}$              & $0.7\,\rm{kpc}$\\
$\Phi_{\rm{disk}}(R,z)$ & = &\parbox{58mm}{\begin{flushright} \begin{equation*}
\label{EqMiyaN}
 -\frac{\rm{GM_{disk}}}{\left(R^2+\left(r_{\rm{disk}}+{\sqrt{(z^2+b^2)}}\right)^2\right)^{1/2}}
%\Phi_{\rm{disk}}(r)=-2{\pi}G\Sigma_{0}{r^2_{\rm{disk}}}\left[\frac{1-\exp^{-r/r_{\rm{disk}}}}{r}\right]
\end{equation*} \end{flushright}}   & $\rm{M}_{disk}$        &  $2.86\times{10^{10}}\,\rm{M}_{\odot}$ & $10.0\times{10^{10}}\,\rm{M}_{\odot}$\\
& & & $\rm{r}_{disk}$        & $5.4\,\rm{kpc}$              & $6.65\,\rm{kpc}$\\ \\
& & & $\rm{b}$                & $0.3\,\rm{kpc}$              & $0.26\,\rm{kpc}$\\
%$\Sigma_{0}$& $4.6\times10^{8}\,\rm{M}_{\odot}\rm{kpc^{-2}}$\\
%$\rho_{\rm{c}}$        & $277.72\;h^2\rm{M_{\odot}}$          & \\
& & & & &\\
\hline
\hline
%$\rm{M}_{bulge}$      & $3.4\times{10^{10}}\,\rm{M}_{\odot}$\\
%$\rm{r}_{bulge}$      & $0.7\,\rm{kpc}$\\
%$\rm{M}_{disk}$       &  $10.0\times{10^{10}}\,\rm{M}_{\odot}$\\
%$\rm{r}_{disk}$       & $6.65\,\rm{kpc}$\\
%$\rm{b}$              & $0.26\,\rm{kpc}$\\
%$\rm{M}_{vir}$        &  $1.5\times{10^{12}}\,\rm{M}_{\odot}$\\
%$\rm{r}_{vir}$        & $294\,\rm{kpc}$\\
%$\rm{d}_{\rm{M31-MW}}$& $779\,\rm{kpc}$\\    
 
\end{tabular}
\caption{Parameters used for the M31 and Milky Way potential. The M31
parameters are consistent with \protect\cite{2006MNRAS.366..996G} and \protect\cite{Widrow2005} and
the Milky Way disk and bulge parameters are taken from
\protect\cite{2005ApJ...635..931B}.}
\label{tabPot}
\end{table*}
 
To assess the stability of satellite galaxy planes through time, the
numerical model focuses on a simulated set of satellite galaxies
around M31. We create 30 satellite galaxies as point masses initially
set on a plane aligned with the disk of M31 ($z=0$), and will be referred to
as an $equatorial$ plane or as the $\theta=0^o$ plane. We construct the
initial $\theta=0^o$ plane by giving the satellites a random value of
their z-position in the range -5 to +5 \kpc\ drawn from an uniform distribution.  
Additionally, their
radial distances are randomly scattered within 50 and 250 kpc of the
centre of M31 potential. Velocities of the satellites are determined by
two components - the velocity on the plane (planar velocity) and
perpendicular to the plane (perpendicular velocity). The planar
velocity is calculated for right-handed co-rotation and restricted to
keep the satellites on elliptical orbits bound to M31. Orbital ellipticities
range from $\epsilon$ = [0.5, 1.0] to create nearly circular orbits. Velocities perpendicular to the
plane are set to 0.0 $\kms$ in order to create a motion that is completely
confined to the plane. This allows us to observe the effects of various
parameter changes more clearly. By rotating this initial plane and its
velocities about a given axis, we later manipulated the plane's
alignment with the M31 galactic disc and dark matter halo equator.  
 
The numerical models are integrated for 5 Gyrs, with snapshots
at intervals of 0.1 Gyrs, using a Leapfrog algorithm \citep{Springel2005}. We begin by looking for a collection of 10
or more satellites in the plane that was set at $t = 0 $ (the initial
plane), that includes the centre of M31. The root-mean-square distance
($D_{\rm{rms}}$) of a satellite galaxy to the plane is taken to be $\leq$15
\kpc\ for it to be considered a part of the planar formation. The
plane considered here is more than twice the thickness of the M31 Vast Thin Plane of Satellites (VTPoS) and about the thickness of the MW Great Polar Plane. The
number of satellites creating a plane is smaller than the number of
satellites in both M31 and MW planes. If we cannot find 10 or more
satellites within 15\kpc\ at a given time-step, it is regarded as a
snapshot where a plane cannot be observed. We search for satellites
that are in the above mentioned range of the initially set plane at
each time-step. We count the number of satellites on the plane and
divide by the number of satellites to calculate the probability of
seeing a satellite on a plane at a given time step- $P$. The
probability is obtained by averaging values from 750 orbital
integrations for each variation of each parameter tested.

\section{Effect of the Milky Way on the Plane of Satellites}\label{MW_pl}
Before exploring the effects of properties such as perpendicular velocity, we investigate the role of the MW on the probability of seeing satellite planes. Throughout the paper tests are conducted with a static potential representing the MW. We also conduct our tests in a M31 potential which gave a smooth dark matter halo, and an environment void of any dark subhalos. Our numerical model (Sec. ~\ref{NumMod}) Milky Way keeps the MW at a distance of 779 kpc and conducts tests for an integration time of 5 Gyrs. The M31 plane is observed to be inclined at $\sim$50$\degree$ to the disk of the galaxy \citep{Ibata2013}. Therefore, we also introduce an incline to our satellite plane formations- where plane orientation and satellites' total velocities are rotated at an angle of $\theta$ about the $x$ axis
from 0 to 90$\degree$ in 15$\degree$ intervals.

\begin{figure}
\includegraphics[width = 1.1\columnwidth]{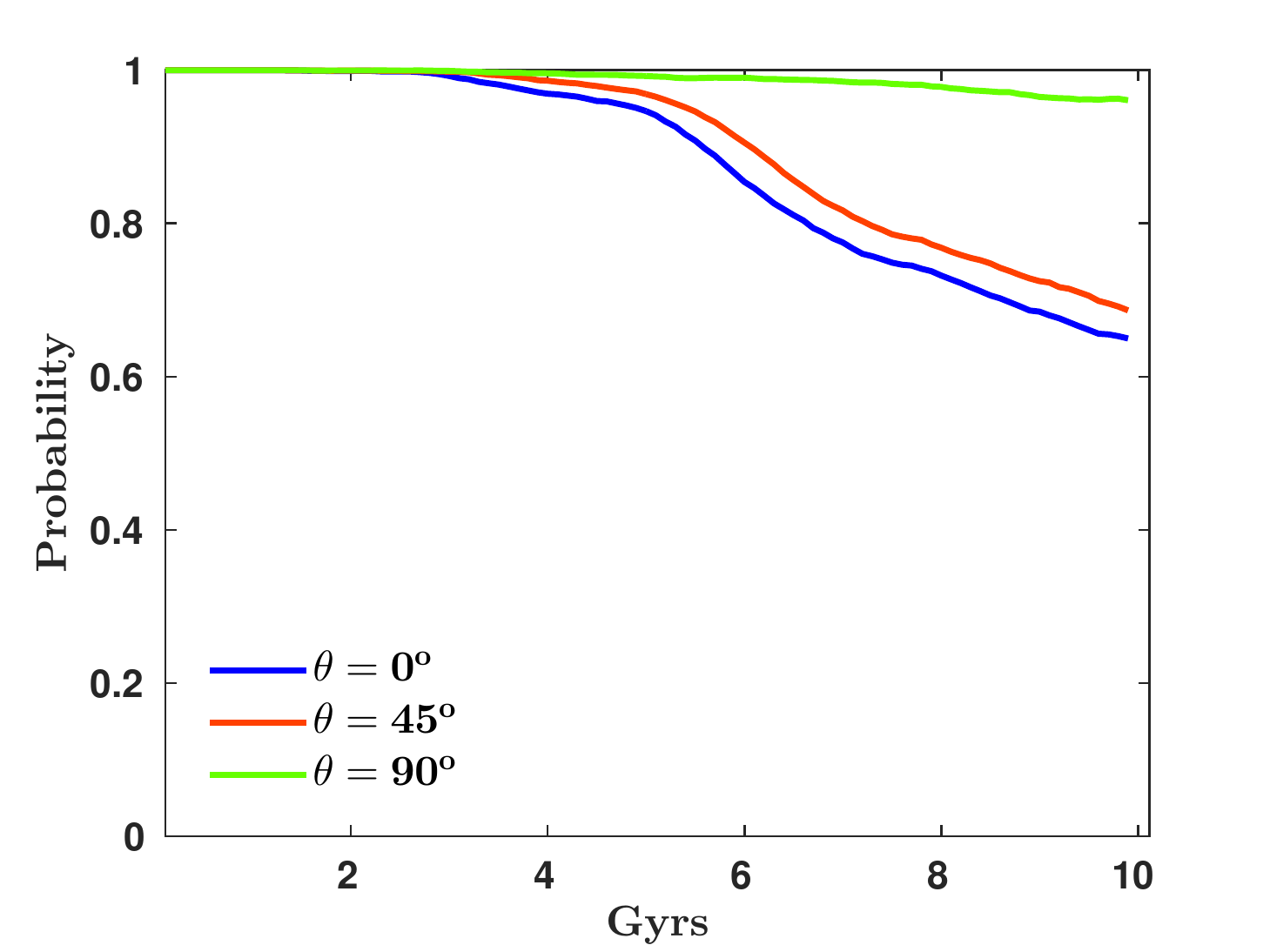}%{Plots/New/Velocity_init_pls1_q3}     
\caption{Probability ($P$) of finding a satellite (of mass $10^{9} \msun$) on planes at varying angles to the $z=0$ plane in the $x$ direction, in a spherical dark-matter halo during 10 Gyrs.}
\label{Fig.Gyrs10}
\end{figure}

The resulting probabilities for a plane lasting through the 5 Gyrs without the destructive forces of additional dark matter halos remain high ($P \geq 0.9$) for planes of all orientations. Extending the integration time to 10 Gyrs, we noticed a striking difference in the probabilities for the existence of the satellite plane (Fig.~\ref{Fig.Gyrs10}). But the equatorial planes show a dramatic drop in $P$ after the 5 Gyr mark. We find a little less than $P = 0.6$ probability for detecting an equatorial plane by the end of 7 Gyrs. Finer examinations of the orbits show that the critical factor in dispersing the equatorial planes is the long-term effect of the MW, positioned at an angle of $\sim$10$\degree$ off the M31 galactic disk. Another important observation is that polar planes are not affected by the MW position in 10 Gyrs of orbital time. With the MW nearly tangent to polar planes of M31 and at a larger distance than for equatorial orientations, its pull on a plane of satellites is minimal. This dispersion of equatorial planes is not seen in simulations without the MW potential. The first 5 Gyr integration period shows no pronounced effects from the MW potential on the plane of M31 satellites. As for most of recent history the MW maintains a distance larger than the current distance to M31 for large portion of the integration time, a plane's probability of survival for an initial 5 Gyr integration period is not greatly affected by the MW.

We can conclude that the MW's influence in the long-term survival of a plane in M31, varies with position and movement. The static Milky Way's influence on a satellite plane in M31 appears after the 5 Gyr period that is examined by the tests and models explored next (Sections~\ref{Stb_VelMass} and ~\ref{Stb_NSH}). Proper motion of the Milky Way puts it at a larger distance through the explored 5 Gyr period into the past. However, the nature of the nearest large neighbour's influence during integration times $\leq$ 10 Gyrs becomes non-negligible and distance to the Milky Way becomes crucial to maintaining the plane of satellites.  

Members of VTPoS share their neighbourhood with larger satellites such as M33, that may act as perturbers of the plane and influence a plane's longevity. While the more massive satellites of M31 (and the higher mass VTPoS members like M32) are likely to have a significant effect on our simulated plane, including such perturbers in our simulation may cloud effects from the more intrinsic properties that we explore in this paper.
%For the 5 Gyr duration of the above runs there are no pronounced effects from the MW potential -kept at a distance of 779 kpc- on the plane of M31 satellites. After extending the integration time to 10 Gyrs, we noticed a striking difference in the probabilities for the existence of the satellite plane (Fig.~\ref{Fig.Gyrs10}). Equatorial (0$\degree$) and polar (90$\degree$) planes have a high probability ($P>0.8$) of lasting through the 5 Gyrs without the destructive forces of additional dark matter halos. But the equatorial planes show a dramatic drop in $P$ after the 5 Gyr mark. We find a very small probability of detecting an equatorial plane by the end of 7 Gyrs. Finer examinations of the orbits show that the critical factor in dispersing the equatorial planes is the MW, positioned at an angle of $\sim$10$\degree$ off the M31 galactic disk. Another important observation is that polar planes are not affected by the MW position in 10 Gyrs of orbital time. With the MW nearly tangent to polar planes of M31, its pull on the plane of satellites is minimal. This dispersion of equatorial planes is not seen in simulations without the MW potential. Therefore, we can conclude that the position and movement of the MW plays a crucial role in the survival of a plane in M31. As for most of recent history the MW maintains a distance larger than the current distance to M31 for large portion of the integration time, the equatorial plane has a higher probability of survival than a polar plane.  

\section{Effects of Velocity and Mass on a Plane of Satellites around M31}\label{Stb_VelMass}
We expect that the velocity and mass of satellite galaxies will play crucial roles in
maintaining a stable planar formation as they are major factors in determining the dispersion and
self-interaction of satellites. In the following section, we will consider the impact  
of these on the longevity of satellite planes, by using the numerical model established in Sec~\ref{NumMod}.  
 
\subsection{Variation of the Velocity Perpendicular to the Plane}\label{StbVel}
This section considers the addition of velocity components perpendicular to the planar orbits of the satellites. Velocities are drawn from a Gaussian distribution with a mean, $\mu$ =
0 $\kms$, and standard deviations, $\sigma$, of 0, 5, 10, 20, 30, 50
and 75 $\kms$ all calculated in the reference frame  centred on M31. The mass of the satellites are
chosen to be $10^9 \msun$, a value representative of the halo population of M31. The satellites are placed initially on the $ y= 0$ plane, in a polar formation with respect to the disk of the M31 potential (Sec. ~\ref{NumMod})\ to discount any effect on the plane from the disk of the host galaxy.
 
Fig.~\ref{Fig.1} shows the effects of velocity on a plane; remember that an initial population of 30 dwarf galaxies were considered and a plane is defined as consisting of 10 satellite systems. Clearly, if there was no off-plane velocity and gravitational interactions between the dwarfs, all members will retain their planar configuration. This is reflected in Fig.~\ref{Fig.1} initially as $P$= 1.0, but after a couple of Gyrs, the probability for detecting a plane of 10 satellites from a population of 30 steadily falls until, after 5 Gyr, the probability is $> 90\%$. As the off-plane velocity is increased, the probability for detecting a plane of 10 satellites drops relatively rapidly.
Examining Fig.~\ref{Fig.1}  in more detail, it is clear that dispersions in the range 5-20 $\kms$ results in steady decline in the observation of a remaining plane from an initial population.
 
Increasing the velocity dispersion significantly reduces the longevity of the planar structure. At 30 $\kms$
the effect is to reduce the observability of satellite planes to 0.5 in the cosmologically short timescale of 1 Gyr, although it reaches a minimum of 0.4 at 0.5 Gyrs. This drop and subsequent rise suggests that the detection of planar structures after 0.5 Gyrs are due to a combination of a smaller number of remaining members on the plane plus interlopers flying through.
 
Increasing the velocity dispersion results in a more rapid demise of planar structures. At the extreme considered here
the plane disperses almost immediately. This has significant implications for the potential accretion origin of the observed planar
structures, something we will return to in Sec.~\ref{Discn}.
 
\subsection{Variation of Satellite Galaxy Masses}\label{StbMass}
 
Simulations of cosmic structure growth produce a broad range of masses for baryonic and dark satellites, as shown in \citet{Sawala2014}, where hundreds of dwarf galaxies in a $10^6$- $10^{12}$ $\msun$ range are found within radius $\leq$ 300 $\kpc$ of a MW-like host. Observationally, the stellar masses for the known satellites of M31 are found in the range $10^6$ - $10^{9}$ $\msun$ \citep{Shaya2013}, and dwarf galaxies (especially dwarf spheroidals) are known to exhibit high mass-to-Light ratios \citep{Mateo2011}, and hence this stellar component is thought to reside in a large dark matter component. To account for this and to test the influence of different satellite masses on the stability of planes of satellites, we give the dwarfs individual masses of $10^7$,$10^8$, $10^9$, $1.5$x$10^9$ and $10^{10}$ $\msun$, encompassing the observed broad range of satellites.

\begin{figure}
\includegraphics[width = 1.1\columnwidth]{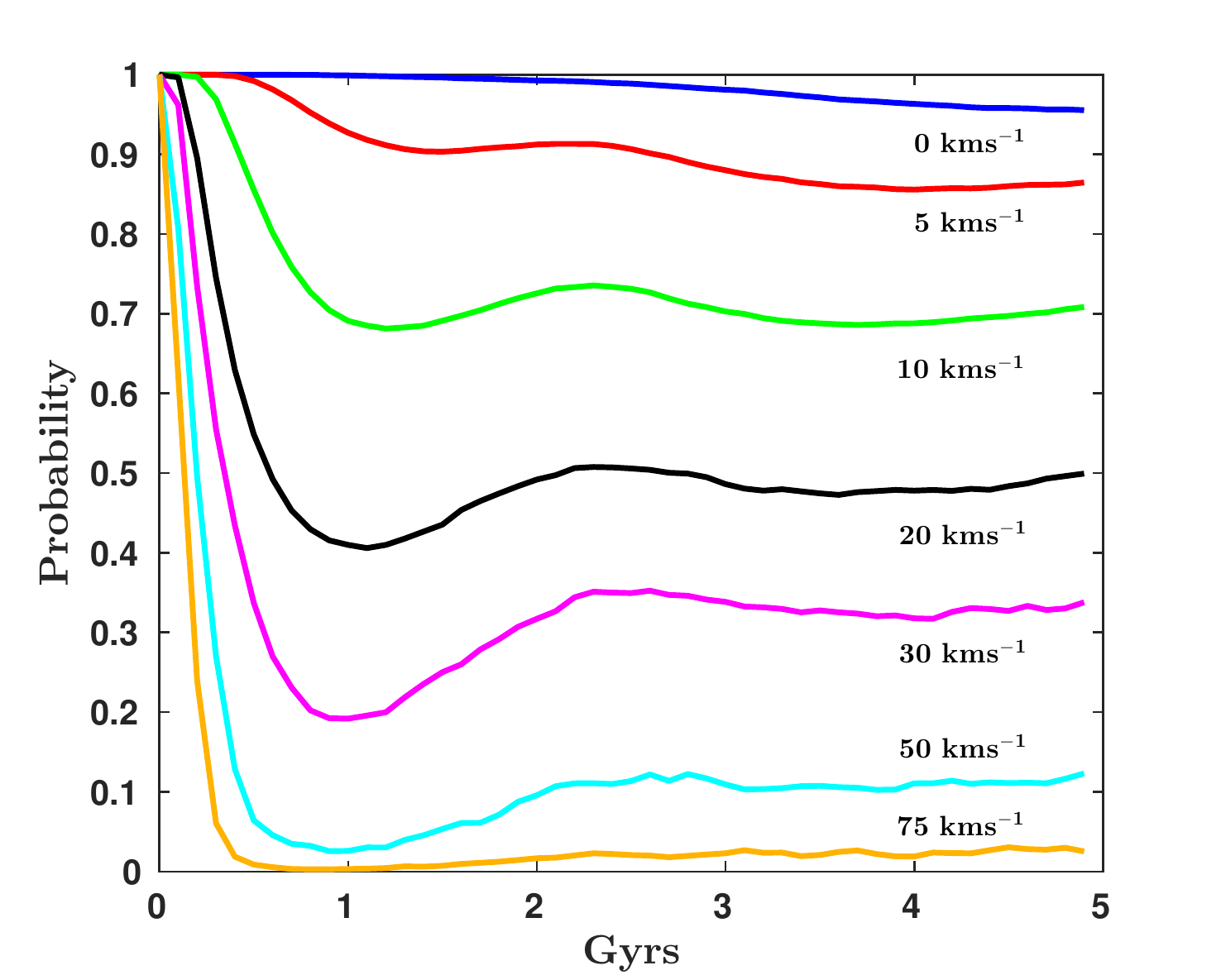}%{Plots/New/Velocity_init_pls1_q3}     
\caption{Probability of finding a satellite on $z=0$ plane ($P$)- varying velocities ($\kms$) perpendicular to plane of satellites of mass  $10^{9} \msun$ }
\label{Fig.1}
\end{figure}

\begin{figure}
\includegraphics[width = 1.1\columnwidth]{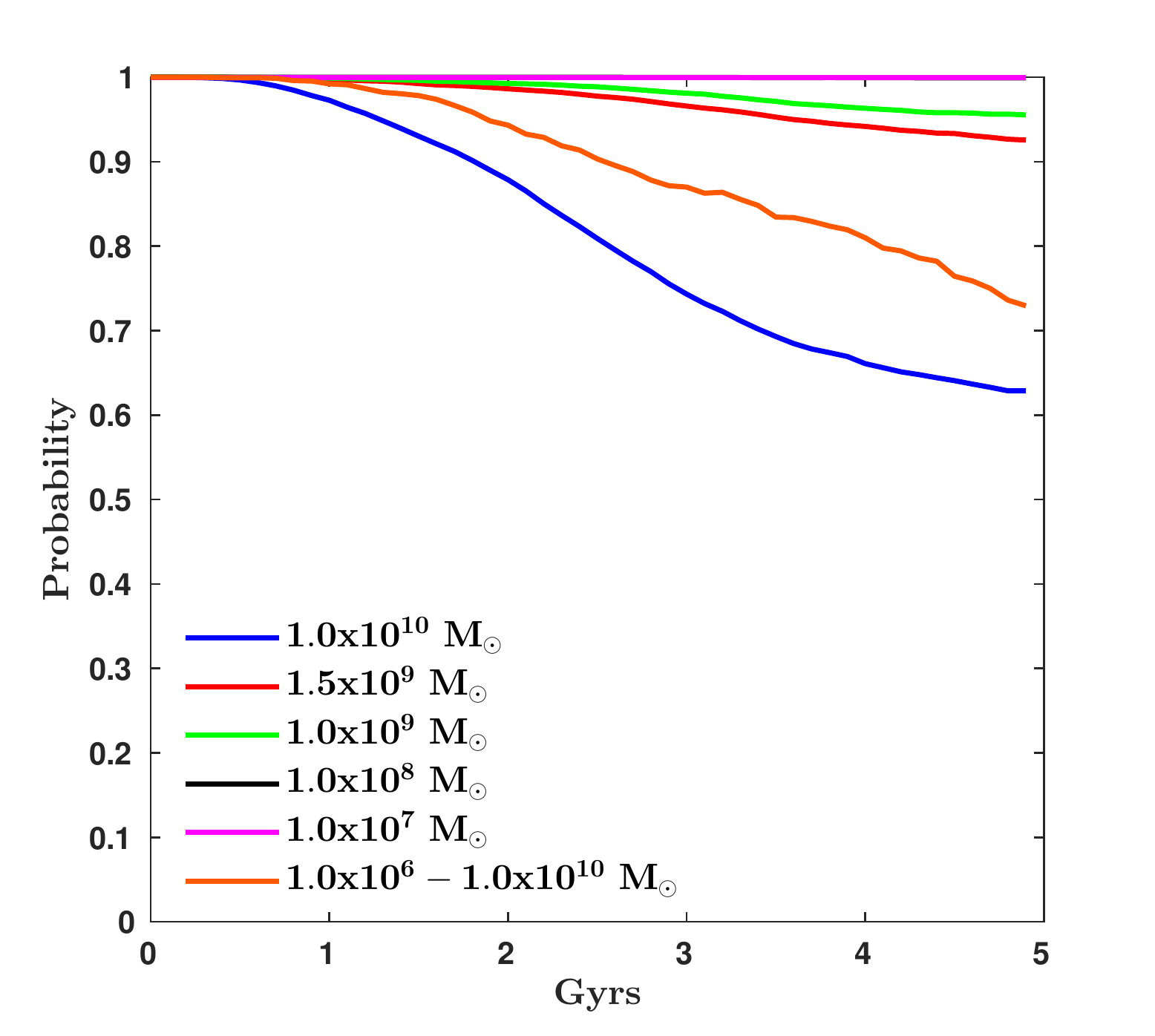}%{Plots/New/Mass_init_pls1_q1_v2}    
\caption{Probability of finding a satellite on $z=0$ plane ($P$)- varying mass of satellites ($\msun$) for satellites with plane perpendicular velocity of $0 \kms$}
\label{Fig.2}
\end{figure}

As we have already seen in Fig.~\ref{Fig.1}, the tangential velocity vector  heavily influences the longevity of satellite planes, so here we consider the extreme case with all of the satellites initially on the $z=0$ plane with a zero perpendicular velocity. The results of the orbital integration in this scenario are presented in  Fig.~\ref{Fig.2}. As expected, the lower mass satellites undergo the weakest self-interactions, with a significant probability for the identification of a plane of satellites after 5 Gyrs. This self-interaction increases as the satellite masses are increased, leading to a more rapid destruction of the initial satellite plane. The orange line represent probability changes for planes where satellite galaxies are given a range of masses from $10^{7}$ to $10^{10}$ $\msun$. Satellites with larger masses affect the smaller satellites to disrupt their orbits on the planes.

It is important to remember that the results presented in Fig.~\ref{Fig.2} are for the most ideal initial conditions with no perpendicular velocity. So comparing Figs~\ref{Fig.1} and~\ref{Fig.2} we can draw interesting conclusions on both the effects of perpendicular velocities and
masses on the longevity of a satellite plane; considering realistic masses for the dwarf satellite population, even in the most idealistic and  
unphysical situation, an extremely cold accretion with no velocity dispersion out of the plane, any plane of satellite galaxies will disperse in
a few Gyrs.  
 
%%%%%%--
 
\section{Influence of Non-Spherical Halos and Relative Orientation of the Plane }\label{Stb_NSH}
So far, we have considered a simple spherical potential to represent dark matter distribution of both M31 and the MW.  
However, cosmological simulations suggest that realistic dark matter halos should be flattened or even triaxial; it is well known
that orbits in non-spherical halo orbits precess and we would expect that such precession will significantly impact the  
longevity of planar structures. \citet{Bowden2013} consider planes in triaxial halos, to show that, through precession,  they double their thickness by the end of 10 Gyrs when the plane is not aligned with the semi-major or semi-minor
axis of the triaxial host halo they inhabit. In this subsection we undertake a broader examination of the
effects of non-spherical halos on a plane of galaxies, by `flattening' the otherwise spherical NFW halo profile.
 
To study the influence of flattened halo potentials, we introduced a flattening parameter, $q$,
 to the $z$ axis of the standard NFW equation (Eq.~\ref{EqNFW}).
For this, we
replace $r^2$ of $\Phi_{\rm{halo}}(r)$ with
\begin{equation}
r^2=x^2 + y^2 + (z*q)^2
\end{equation}
We consider flatness parameters of $q$ =1.67, 1.43, 1.25, 1.11, 1.0, 0.975, 0.95, 0.9, 0.8, 0.7 and 0.6. $q < $ 1.0 produces prolate halos while $q > 1.0$ produces oblate halos (with $q$ =1.11 being equivalent to $q=$1/0.9 etc.).  
Again, we restrict the satellites to an equatorial plane of $z$=0, each given a
perpendicular velocity of 0 $\kms$, with a mass of  $10^{9}$ $\msun$ each; examining Fig.~\ref{Fig.2}
shows that this mass scale and lower show little notable gravitational self-interaction between the satellites.
 
\subsection{Inclined Planes in Non-Spherical Halos}\label{Incl_Pl}
 
%\color{red}As observed in the results of Sec.~\ref{MW_pl}, M31 plane's orientation to the host's stellar disc may hold a prominent position in list of influencers of plane stability. Within the MW, the corresponding plane of satellites is also seen to be perpendicular to the plane \citep{Kroupa2005}. \color{black} Most MW-M31 size galaxies in cosmological simulations show their disk to be aligned with their dark matter halo \citep{Vera-Ciro2011}.  It is therefore likely that the M31 VTPoS is misaligned with the axes of its host's major dark matter halo.  

As described in the results of Sec.~\ref{MW_pl}, we found that the satellite planes's orientation to the host halo plays a key role on the plane's stability. In the case of the Milky Way, the observed plane of satellites is perpendicular to the stellar disk \citep{Kroupa2005} whereas in the case of M31 this is not the case. Most MW-M31 size galaxies in cosmological simulations show their disk to be aligned with their dark matter halo \citep{Vera-Ciro2011}. It is therefore likely that the M31 VTPoS is misaligned with the axes of its host's major dark matter halo.
 
\begin{figure*}% {first intro of inclines}
\centering
%  \subfigure[$q$=1.0]{\includegraphics[width=0.45\textwidth]{Plots/qplots/T7_pls1_q1.png}\label{Fig.4.a}}
%  \subfigure[$q$=0.975]{\includegraphics[width=0.45\textwidth]{Plots/qplots/T7_pls1_q2.png}\label{Fig.4.b}}
%  \subfigure[$q$=0.95]{\includegraphics[width=0.45\textwidth]{Plots/qplots/T7_pls1_q3.png}\label{Fig.4.c}}
%  \subfigure[$q$=0.9]{\includegraphics[width=0.45\textwidth]{Plots/qplots/T7_pls1_q4.png}\label{Fig.4.d}}
%  \subfigure[$q$=0.8]{\includegraphics[width=0.45\textwidth]{Plots/qplots/T7_pls1_q5.png}\label{Fig.4.e}}
%  \subfigure[$q$=0.7]{\includegraphics[width=0.45\textwidth]{Plots/qplots/T7_pls1_q6.png}\label{Fig.4.f}}
\includegraphics[width = 0.8\textwidth ]{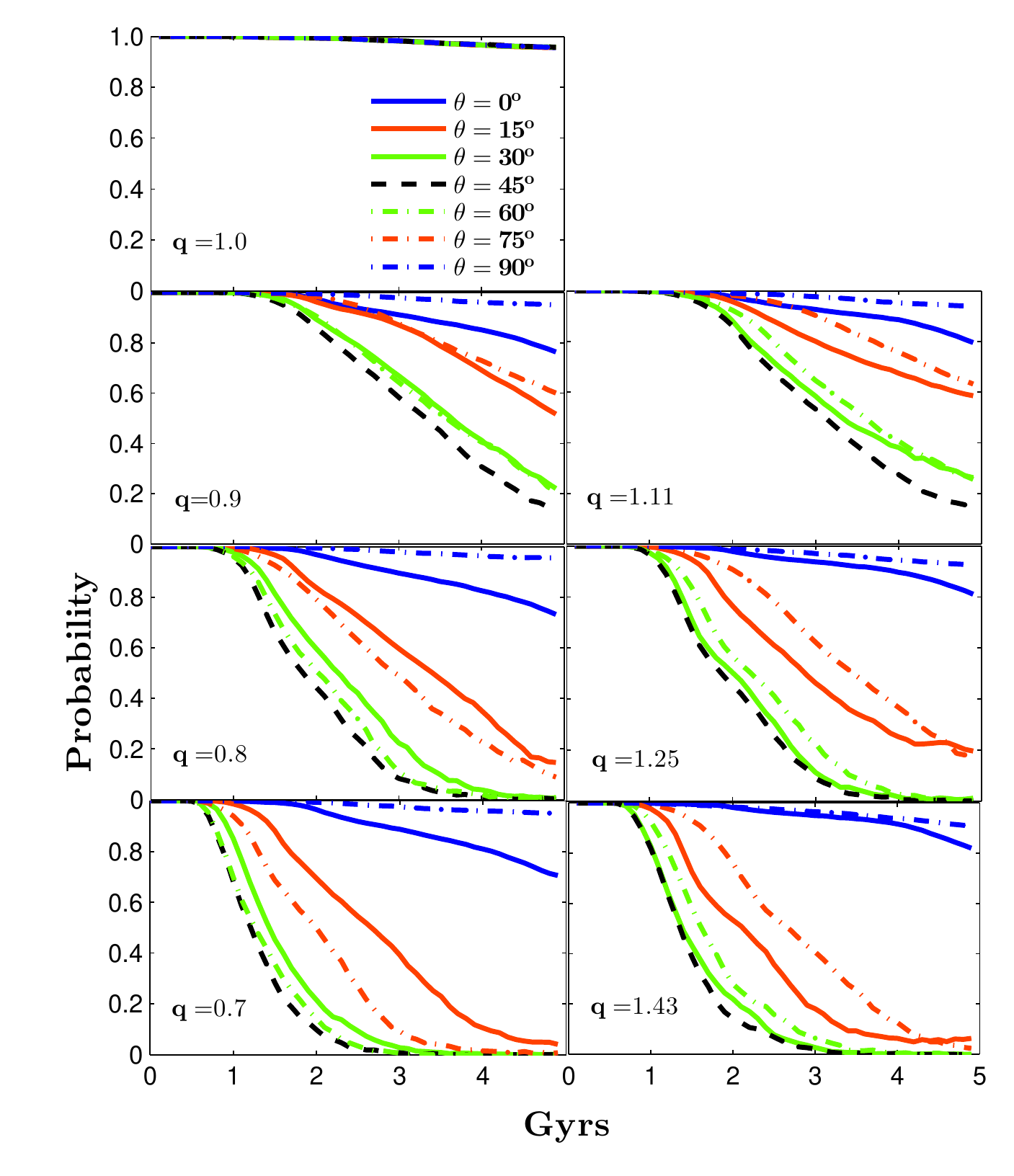}%{Plots/New/Probs/All_probs5}  
%  \subfigure[$q$=1.0]{\includegraphics[width=0.47\textwidth]{Plots/New/Probs/Probs_init_pls1_q1_v2.eps}\label{Fig.3.a}}
%  \subfigure[$q$=0.975]{\includegraphics[width=0.45\textwidth]{Plots/New/Probs/Probs_init_pls1_q2.eps}\label{Fig.3.b}}
%  \subfigure[$q$=0.95]{\includegraphics[width=0.45\textwidth]{Plots/New/Probs/Probs_init_pls1_q3.eps}\label{Fig.3.c}}
%  \subfigure[$q$=0.9]{\includegraphics[width=0.45\textwidth]{Plots/New/Probs/Probs_init_pls1_q4.eps}\label{Fig.3.d}}
%  \subfigure[$q$=0.8]{\includegraphics[width=0.45\textwidth]{Plots/New/Probs/Probs_init_pls1_q5.eps}\label{Fig.3.e}}
%  \subfigure[$q$=0.7]{\includegraphics[width=0.45\textwidth]{Plots/New/Probs/Probs_init_pls2_q6.eps}\label{Fig.3.f}}
\caption{Probability of finding a satellite on an inclined plane ($P$) of 10 or more galaxies within $D_{\rm{rms}}$ $\leq$ 15, and of mass $10^9$ $\msun$ - varying  $'flatness'(q)$ of the M31 halo}
\label{Fig.3}
\end{figure*}

Here, we introduce various inclinations of the plane of satellites
with respect to flattened halos with different values of $q$. Initially the satellite galaxies are on a z=0 equatorial plane (with
velocities restricted to the plane so the off plane dispersion is zero). Following the method of Sec.\ref{MW_pl}, the plane of satellites and their
velocities are rotated at an angle of $\theta$ about the $x$ axis
from 0 to 90$\degree$ in 15$\degree$ intervals. The orbits are
integrated for 5 Gyrs and the values of the probability are shown in Fig.~\ref{Fig.3}.
 
\begin{table*}  
 \centering
 
 \begin{minipage}{140mm}%{1.0\textwidth}%{140mm}
%  \resizebox{\columnwidth}{!}{%
  \caption{Probability of seeing satellites on planes ($P$)- System of
30 satellites with 0 $\kms$ perpendicular velocity, satellite mass =
$10^9 \msun$, minimum number of satellites on a plane = 10 $\kpc$ and
$D_{\rm{rms}}$ = $\pm$ 15 $\kpc$ } \label{Table.2}
  \begin{tabular}{|p {17mm} | p {8mm}| p {8mm}| p {8mm}| p {8mm}| p {8mm}| p {8mm} |p {8mm} |p {8mm} |p {8mm}|p {8mm}|p {8mm}|}
\hline
%      &    Mass &    satellites & cyan. vel &    halos &    min plane sat & max dist &      \\
% \hline
%  & $10^9 \msun$ & 30 &    0 $\kms$ &     0 $\kms$ &      10 $\kpc$  & $\pm$15 $\kpc$ & \\      
% \hline
 \multicolumn{12}{|c|}{\textbf{Part (a).} Average probability of seeing a satellite on initial planes during 5 Gyrs}\\
 \hline
  \backslashbox{Angle}{q}&1.67&1.43&1.25&1.11&1.0&0.975&0.95&0.9&0.8&0.7&0.6\\  
\hline
0$\degree$&0.99&0.99&0.98&0.98&0.93&0.93&0.93&0.92&0.91&0.90&0.89\\
15$\degree$&0.37&0.49&0.69&0.91&0.94&0.93&0.92&0.86&0.68&0.54&0.43\\
30$\degree$&0.23&0.31&0.46&0.76&0.95&0.94&0.89&0.73&0.48&0.32&0.25\\
45$\degree$&0.23&0.29&0.42&0.71&0.96&0.94&0.88&0.67&0.40&0.27&0.21\\
60$\degree$&0.28&0.35&0.49&0.77&0.97&0.96&0.91&0.72&0.43&0.28&0.21\\
75$\degree$&0.49&0.59&0.74&0.92&0.98&0.98&0.96&0.88&0.62&0.41&0.30\\
90$\degree$&0.97&0.97&0.98&0.98&0.98&0.98&0.98&0.98&0.98&0.98&0.98\\
\hline
 \multicolumn{12}{|c|}{\textbf{Part (b).} Average probability of seeing a satellite on best- fit planes during 5 Gyrs }\\
\hline  
  \backslashbox{Angle}{q}&1.67&1.43&1.25&1.11&1.0&0.975&0.95&0.9&0.8&0.7&0.6\\  
\hline   
0$\degree$&0.99&0.99&0.99&0.96&0.96&0.95&0.95&0.94&0.93&0.98&0.98\\
15$\degree$&0.71&0.75&0.82&0.94&0.96&0.96&0.94&0.90&0.79&0.70&0.62\\
30$\degree$&0.55&0.62&0.71&0.84&0.97&0.96&0.92&0.82&0.67&0.55&0.42\\
45$\degree$&0.48&0.57&0.67&0.81&0.97&0.96&0.91&0.80&0.62&0.48&0.38\\
60$\degree$&0.49&0.59&0.69&0.84&0.98&0.97&0.93&0.81&0.63&0.45&0.34\\
75$\degree$&0.62&0.69&0.80&0.93&0.98&0.98&0.97&0.90&0.70&0.48&0.33\\
90$\degree$&0.97&0.97&0.98&0.98&0.98&0.98&0.98&0.98&0.98&0.98&0.98\\  
\hline
\end{tabular}
\end{minipage}
%\label{Table.1}
\end{table*}
 
%Init
%0.98     0.98     0.98     0.98     0.98     0.98     0.98
%0.98     0.98     0.98     0.97     0.98     0.98     0.98
%0.98     0.974     0.94     0.92     0.94     0.97     0.98  
%0.98     0.91     0.76     0.71     0.77     0.92     0.98  
%0.98     0.69     0.46     0.42     0.49     0.74     0.98  
%0.99     0.49     0.31     0.29     0.35     0.59     0.97  
%0.99     0.37          0.23     0.23     0.28     0.49     0.97
 
%Fit
%0.99     0.98     0.987     0.98     0.98     0.98     0.99  
%0.99     0.98     0.98     0.98     0.98     0.98     0.99
%0.99     0.98     0.96     0.94     0.96     0.98     0.99
%0.99     0.94     0.84     0.81     0.84     0.93     0.98  
%0.99     0.82     0.71     0.67     0.69     0.80     0.98
%0.99     0.75     0.62     0.57     0.59     0.69     0.97  
%0.99     0.71     0.55     0.48     0.49     0.62     0.97  

Fig.~\ref{Fig.3} presents the survivability of planes (of satellites) at the differing inclinations and  
flattening values, $q$.
Considering equatorial planes ($\theta$ = $0\degree$), we can see that the non-spherical nature of the halo  
has a mild influence on the probability of finding a satellite on a plane.
As the flattening of the potential is increased/decreased from 1.0, the probability for finding a planar distribution experiences little variation with respect to that seen for a fully spherical halo.  
 
Such planes, therefore, possess long-term dynamic stability,
with a high probability of seeing a plane at any given time within 5
Gyrs. As the angle is increased from 45$\degree$- 90$\degree$ the plane
gets closer to the $z$ axis and $P$ is notably increased.
 
The average probability of finding a plane (minimum of 10 satellites)
for 5 Gyrs is given in part (a) of Table~\ref{Table.2}, columns varying halo flatness and
rows varying plane inclination. Their pattern reflect the overall trends
of Fig.~\ref{Fig.3}- average probability changes with incline and $q$
values. It is evident that in extremely prolate ($q$= 0.6) and oblate ($q$=1.67) halos, even if
satellites start on a plane, each satellite has only a probability of
0.21 of being seen on that plane through 5 Gyrs. The potential change
created by a flattened halo does not greatly affect satellites set at $z=0$, allowing those planes to last longer than inclined
planes.
%might want to tell that the MW's
%presence is showing itself in the numbers even now - that's why the
%theta =0 has a lower
 
Planes rotated along the $x$ axis show different probabilities for all
non-spherical halos starting from $q= 0.95$ (Table.~\ref{Table.2}). Though
it appears as a very small change from $q=1.0$, it has a visible
effect on the number of satellites that remain on the initial planar
region for 5 Gyrs. As the rotated planes get closer
to another axis ($z$ in this case), the probability values get higher.
Reducing $q$ to 0.95 will give satellites on any inclined plane a
final probability $\leq$ 0.5 (Fig.~\ref{Fig.3}).  
 
However, starting from $q$ = 0.9 (Fig.~\ref{Fig.3}) the $P$ values decline at a very rapid rate - most clearly seen at the higher angles of inclination. For a plane with a higher inclination about the $x$ axis the probability of seeing a satellite on the initial plane region
decreases. For a 45$\degree$ inclination, the
probability $P$ decreases by $\sim$40\% and the planes disintegrate
faster as the initial plane incline is increased. This trend continues
for other $q \neq 1.0$ values. In fact, the effect is greatly
amplified as $q$ values decrease to form prolate halos or increase to form oblate halos.
In particular for a plane with a 45$\degree$ inclination, the probability
$P$ is similar to that of a random distribution after 5 Gyrs. As we
further reduce $q$, all inclined planes will disintegrate rapidly and reach a random distribution after around 2.5 Gyrs
(Fig.~\ref{Fig.3}). For $q$ = 0.9, planes with an inclination larger than
30$\degree$ to a principal axis are reduced to a probability less than
0.5 in 5 Gyrs. The time scale to reach similar probability for $q$ =
0.8 is effectively halved from 5 Gyrs to 2$\sim$\ 3 Gyrs for all
inclined planes. Further decreasing $q$ will cause the probabilities
to decrease to $P$=0.2 an entire Gyr earlier for all inclined planes,
reducing the lifetime of a visible satellite plane to less than 3
Gyrs. The trends are similar when increasing the oblate nature of the host halo ($q > 1.0$), with the changes to the probability exhibiting a symmetry $q$. %The only notable difference seen for the different types of halos shapes is that $\theta =0\degree$ planes show a final $P$ of ~ 0.8 at 5 Gyrs for prolate halos and only a small decrease from 1.0 for oblate halos.\color{black} %********************
 
The average time a satellite spends on a plane is calculated and given
in part (a) of Table.~\ref{Table.3}. This does not show the average
continual time, but cumulates all the time-steps when a given
satellite is on its initial plane range, and gives the number of years
that it can be spotted in its starting planar formation. Planes set on
the $x$ and $z$ axes are likely to stay for more than 4.4 out of the 5
Gyrs on the plane despite changes of $q$. Inclined planes
clearly reflect the trends of Table~\ref{Table.2} and Fig.~\ref{Fig.3}. For halos with a flatness of
$q=0.8$ planes at 45$\degree$ will be observable on average 2 Gyrs.
But for more prolate halos ($q=0.6$) and more oblate shapes ($q=1.67$), the overall lifetime of all
inclined planes is smaller than 2 Gyrs. Dark matter only simulations
calculate triaxial axis ratios to be around $b/a \sim$ 0.6 and $b/c
\sim $0.4 \citep{Jing2002, Allgood2006}. Therefore, although $q=0.6 $/$1.67$ represents a smaller flattening than the average values for host halos, planes with inclination as small as 15$\degree$ to major axes  have an average lifetime of only 2 Gyrs or less. %As the potential difference changes significantly through the cross-section of an inclined plane, the variations cause the satellites to stray out of their pre-set orbits.
 
\subsection{Fitted Planes and Plane Precession}\label{Fit_Prec}
In the previous sections, we have calculated the probability of finding a satellite
galaxy on the `original' plane it was set on. Flattened halos have shown interesting properties in simulations: the orbits
of satellites display precession through a period of time.
\citet{Ibata2001} demonstrate how the Milky Way is unlikely to have a halo
flatter than $q$ =0.7 with their analysis and simulations of the
Sagittarius dwarf galaxy and its tidal stream. \citet{Bowden2013}\
suggest that precessing planes could not be found for the 30 most massive
satellites examined in their zoom-in dark matter
simulations run for 5 Gyrs. For their test halos, starting from infall
times, the plane found at each timestep was not a slight precession
from the previous plane but an altogether different plane.
 
What if satellite planes change their orientation and precess with
time? As a plane disperses we still may see planar formations at
different angles of inclination. A plane-fitting method should detect
these precessing planes of satellites as the `best-fit plane' for each
time step. For precessing planes, the probability of finding a
satellite on a fitted plane should be larger than the probability of
finding a satellite on the initial plane (calculated in part (b) of
Table~\ref{Table.2}). To test this hypothesis, we used least squares fitting
to find the best fit plane that also contains the center of M31. Here,
a plane retains the same definition of `10 or more satellites
within a distance of $D_{\rm{rms}}$ = 15 $\kpc$ to the best fit
plane'. Plane fitting is applied to each snapshot of the 5 Gyr
integration which are taken at 0.1 Gyr intervals.  
 
The average probability of a satellite being on the best-fit plane at
any given time through 5 Gyrs is given in part (b) of
Table~\ref{Table.2}. The trend of increasing probability as planes move towards the axes from 45$\degree$ and decreased probability
for flatter halos are already apparent in averaged values in an
initial plane (Table~\ref{Table.2} part (a)) and continue here too.
Halos with $ 0.8 \leq q \leq 1.0$ show around 10\% increase of $P$
from the initial plane probability to best-fit probability and more
prolate/oblate shapes have $P$ increased by $\sim$20\%. Notable changes are
only seen for planes not-aligned with the axes. For satellites on
equatorial and polar planes in halos with $q=0.6$ oblateness, we can
see only a 10\% increase between the probability of finding it on its
original plane to being aligned with the best-fit plane of the
snapshot.
 
Part (b) of Table~\ref{Table.3} calculates the average time we can see
a satellite on the best-fit planar formation during 5 Gyrs. We see
that the polar and equatorial satellites spend nearly all the 5 Gyrs
in a planar formation that is larger than 10 satellites and has a
thickness of $\sim$30 $\kpc$. Inclined planes of halos with $ 0.8 \leq q \leq 1.0$ show that satellites stay in the planes at most 4 Gyrs and at least around 2 Gyrs. We can see that a group of satellites that
start out on a plane stay in a planar formation for at least 2 out of
5 Gyrs from its time of assembly, even when orbiting dark matter halos
as prolate as $q= 0.6$ and as oblate as $q=1.67$. However, given that our satellites are confined to the plane due to their initial velocities, we can expect the
satellites with non-zero perpendicular velocities to show a lower
average time on their initial planes. It is important to note here
that we are focusing on planes containing at least 30\% of the entire
satellite galaxy population - it is likely that planar formations with
a smaller number of satellites have a higher probability of being
found, but their significance with respect to the rest of the
population is smaller than what we can see in both observational
examples of M31 and MW.  
 
\begin{table*}
 \centering
 \begin{minipage}{140mm}
  \caption{Average time spent on planes - system of 30 satellites with
0 $\kms$ perpendicular velocity, satellite mass = $10^9 \msun$,
minimum number of satellites on a plane = 10 $\kpc$ and $D_{\rm{rms}}$
= $\pm$15 $\kpc$ } \label{Table.3}
  %\begin{tabular*}{\textwidth}{|c @{\extracolsep{\fill}}|c |c |c |c |c |c |c |}
  \begin{tabular}{|p {17mm} | p {8mm}| p {8mm}| p {8mm}| p {8mm}| p {8mm}| p {8mm} |p {8mm} |p {8mm} |p {8mm}|p {8mm}|p {8mm}|}
  \hline
%     &    Mass &    satellites & cyan. vel &    halos &    min plane sat & max dist &      \\
%  \hline
%  & $10^9 \msun$ & 30 &    0 $\kms$ &     0 $\kms$ &      10 $\kpc$  & $\pm$15 $\kpc$ & \\      
%  \hline
  \multicolumn{12}{|c|}{\textbf{Part (a).} Average time (Gyrs) a satellite spends on its initial plane in 5 Gyrs }\\
  \hline
  \backslashbox{Angle~}{q~}&1.67&1.43&1.25&1.11&    1.0 &    0.975&   0.95 &        0.9 &        0.8 &         0.7 &          0.6 \\  
\hline
 0$\degree$ &4.9&4.9&4.9&4.9& 4.7 &    4.7 &    4.6 &    4.6 &    4.6 &    4.5 &    4.4 \\
 15$\degree$ &1.9&2.5&3.4&4.6& 4.7 &    4.6 &    4.6 &    4.3 &    3.4 &    2.7 &    2.0 \\
 30$\degree$ &1.2&1.5&2.3&3.8& 4.7 &    4.7 &    4.4 &    3.6 &    2.4 &    1.6 &    1.2 \\
 45$\degree$ &1.2&1.4&2.1&3.6& 4.8 &    4.7 &    4.4 &    3.3 &    2.0 &    1.3 &    1.0 \\
 60$\degree$ &1.4&1.7&2.5&3.9& 4.8 &    4.8 &    4.5 &    3.6 &    2.1 &    1.4 &    1.0 \\
 75$\degree$ &2.4&2.9&3.7&4.6& 4.9 &    4.9 &    4.8 &    4.4 &    3.1 &    2.0 &    1.5 \\
 90$\degree$ &4.8&4.9&4.9&4.9& 4.9 &     4.9 &     4.9 &    4.9 &     4.9 &     4.9 &     4.8\\
\hline
\multicolumn{12}{|c|}{\textbf{Part (b).} Average time (Gyrs) a satellite spends on the best-fit plane in 5 Gyrs }\\
 \hline  
  \backslashbox{Angle}{q}& 1.67&1.43&1.25&1.11&    1.0 &    0.975&   0.95 &     0.9 &        0.8 &         0.7 &          0.6 \\  
\hline
 0$\degree$ &4.9&4.9&4.9&4.9& 4.8 &    4.8 &    4.8 &    4.7 &    4.7 &    4.6 &    4.6 \\
 15$\degree$ &3.6&3.8&4.1&4.7& 4.8 &    4.8 &    4.7 &    4.5 &    3.9 &    3.5 &    3.1 \\
 30$\degree$ &2.8&3.1&3.5&4.2& 4.8 &    4.8 &    4.6 &    4.1 &    3.3 &    2.7 &    2.2 \\
 45$\degree$ &2.4&2.8&3.4&4.0& 4.8 &    4.8 &    4.5 &    3.9 &    3.1 &    2.4 &    1.9 \\
 60$\degree$ &2.5&2.9&3.5&4.2& 4.9 &    4.9 &    4.6 &    4.1 &    3.1 &    2.2&    1.7 \\
 75$\degree$ &3.1&3.5&4.0&4.7& 4.9 &    4.9 &    4.8 &    4.5 &    3.5 &    2.4&    1.6 \\
 90$\degree$ &4.9&4.9&4.9&4.9& 4.9 &     4.9 &     4.9 &     4.9 &     4.9 &     4.9 &    4.9\\
\hline
\end{tabular}
\end{minipage}
%\label{Table.2}
\end{table*}
% Init
%4.9    4.9     4.9     4.9    4.9     4.9     4.9  
%4.9     4.9     4.9     4.9     4.9    4.9    4.9
%4.9     4.9    4.7     4.6     4.7     4.9     4.9
%4.9    4.6    3.8     3.6     3.9     4.6     4.9
%4.9    3.4    2.3    2.1    2.5     3.7     4.9
%4.9    2.5    1.5     1.4     1.7     2.9     4.9
%4.9    1.9     1.2     1.1     1.4     2.4     4.8
 
%Fit
%4.9    4.9     4.9     4.9     4.9     4.9     4.9  
%4.9    4.9     4.9     4.9    4.9     4.9     4.9
%4.9     4.9     4.8     4.7    4.8     4.9     4.9
%4.9     4.7     4.2     4.0    4.2     4.7     4.9
%4.9     4.1     3.5     3.4    3.5     4.0     4.9
%4.9     3.8     3.1     2.8    2.9    3.5    4.9
%4.9     3.6     2.8     2.4    2.5    3.1    4.9

%increase of thickness of the plane shown as another proof for there not being a big difference between fitted and original planes..  
 
\begin{figure*}% {normals of fitted planes} % Put Colour Bar !!!!
\centering
  \subfigure[$\theta$=15$\degree$]{\includegraphics[width=0.45\textwidth]{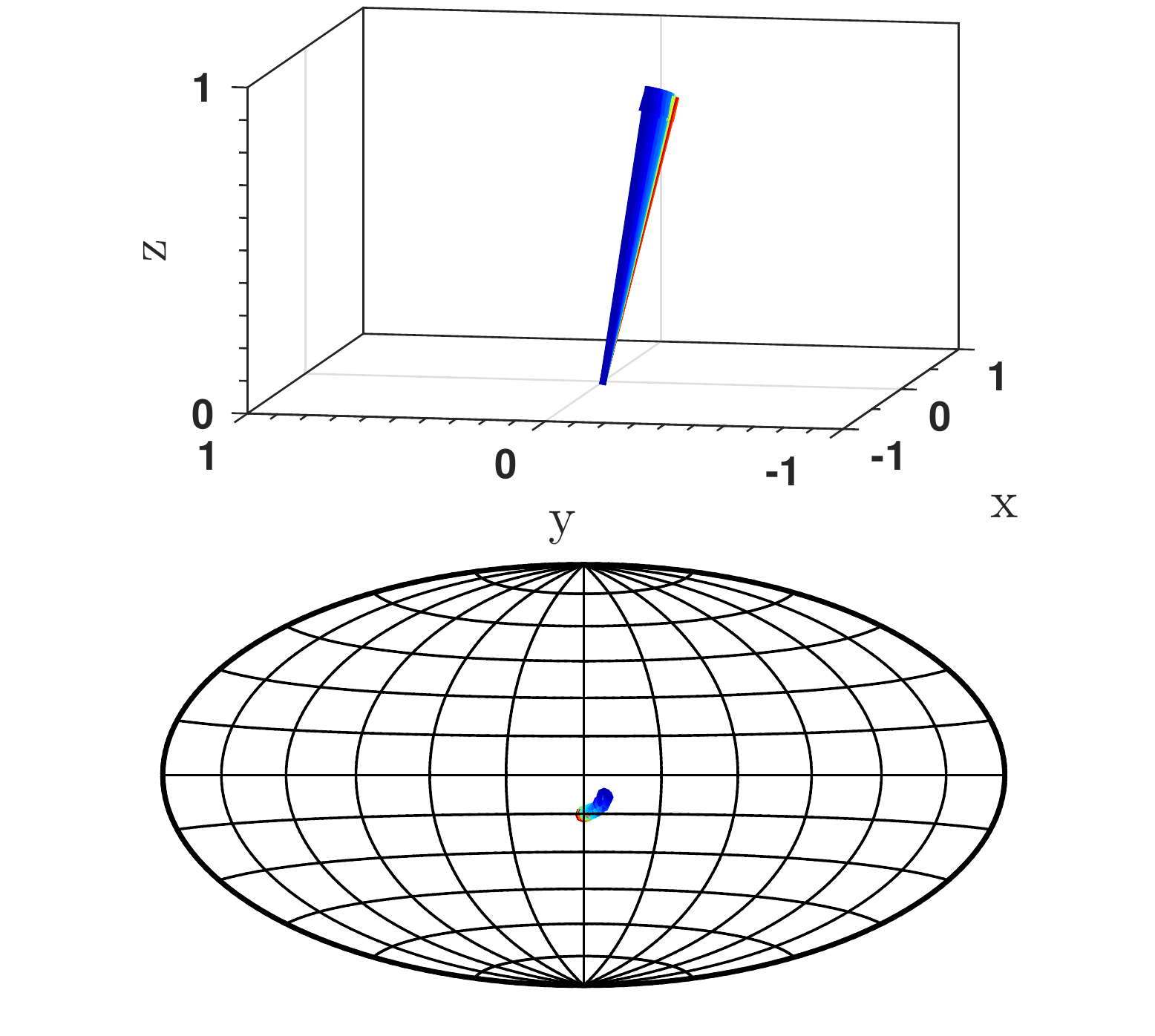}\label{Fig.4.a}}
  \subfigure[$\theta$=30$\degree$]{\includegraphics[width=0.45\textwidth]{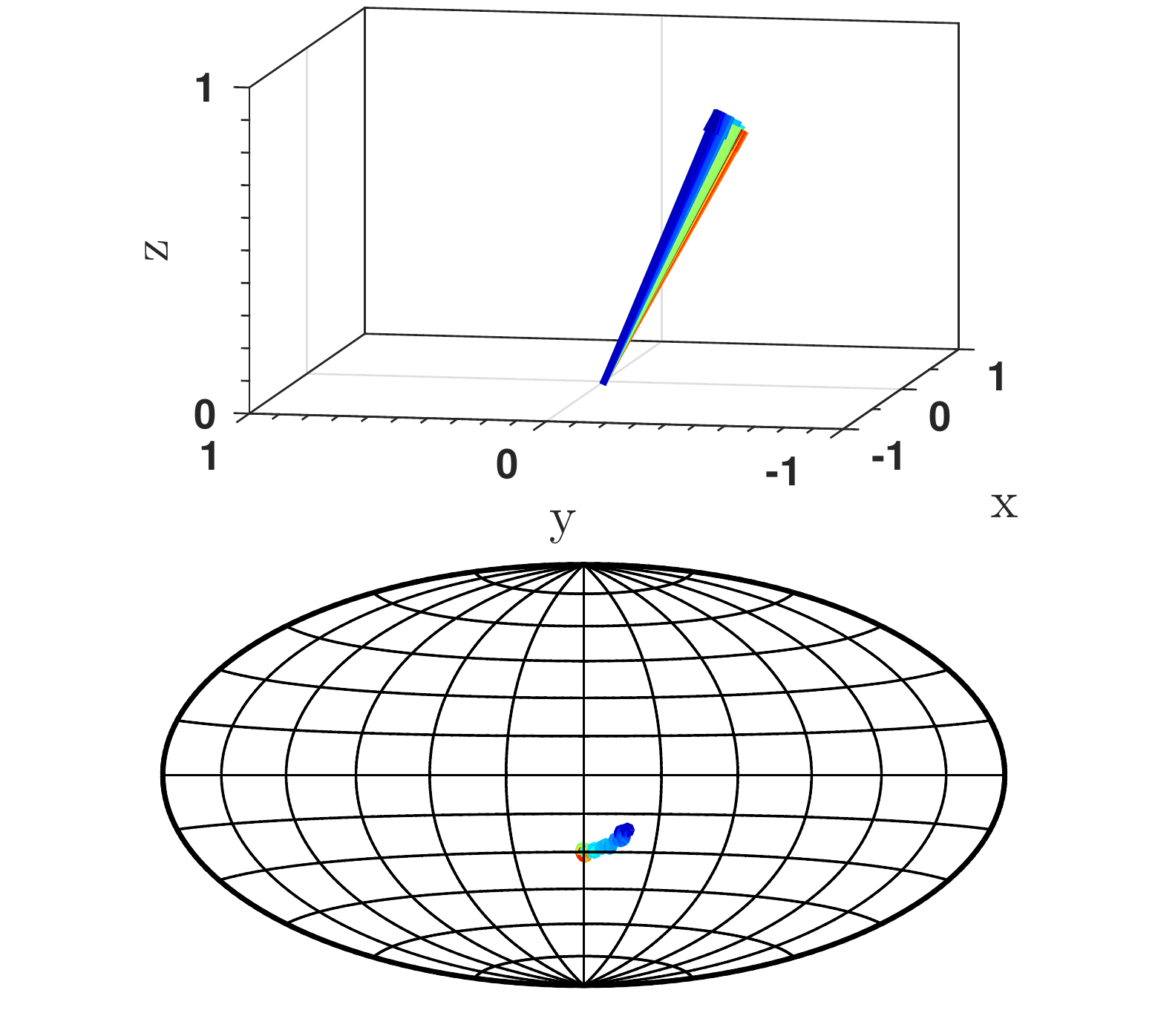}\label{Fig.4.b}}
  \subfigure[$\theta$=45$\degree$]{\includegraphics[width=0.45\textwidth]{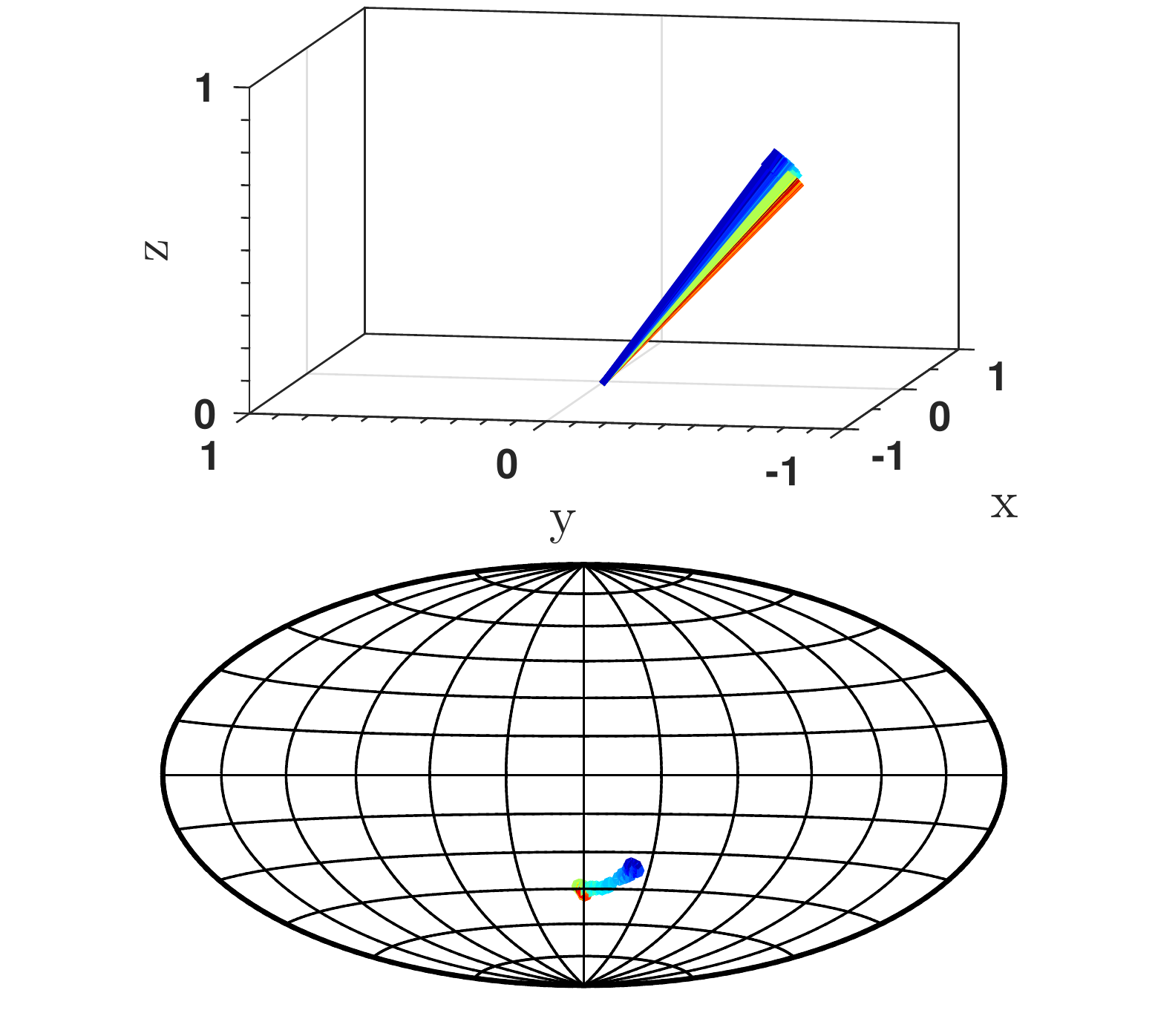}\label{Fig.4.c}}
  \subfigure[$\theta$=60$\degree$]{\includegraphics[width=0.45\textwidth]{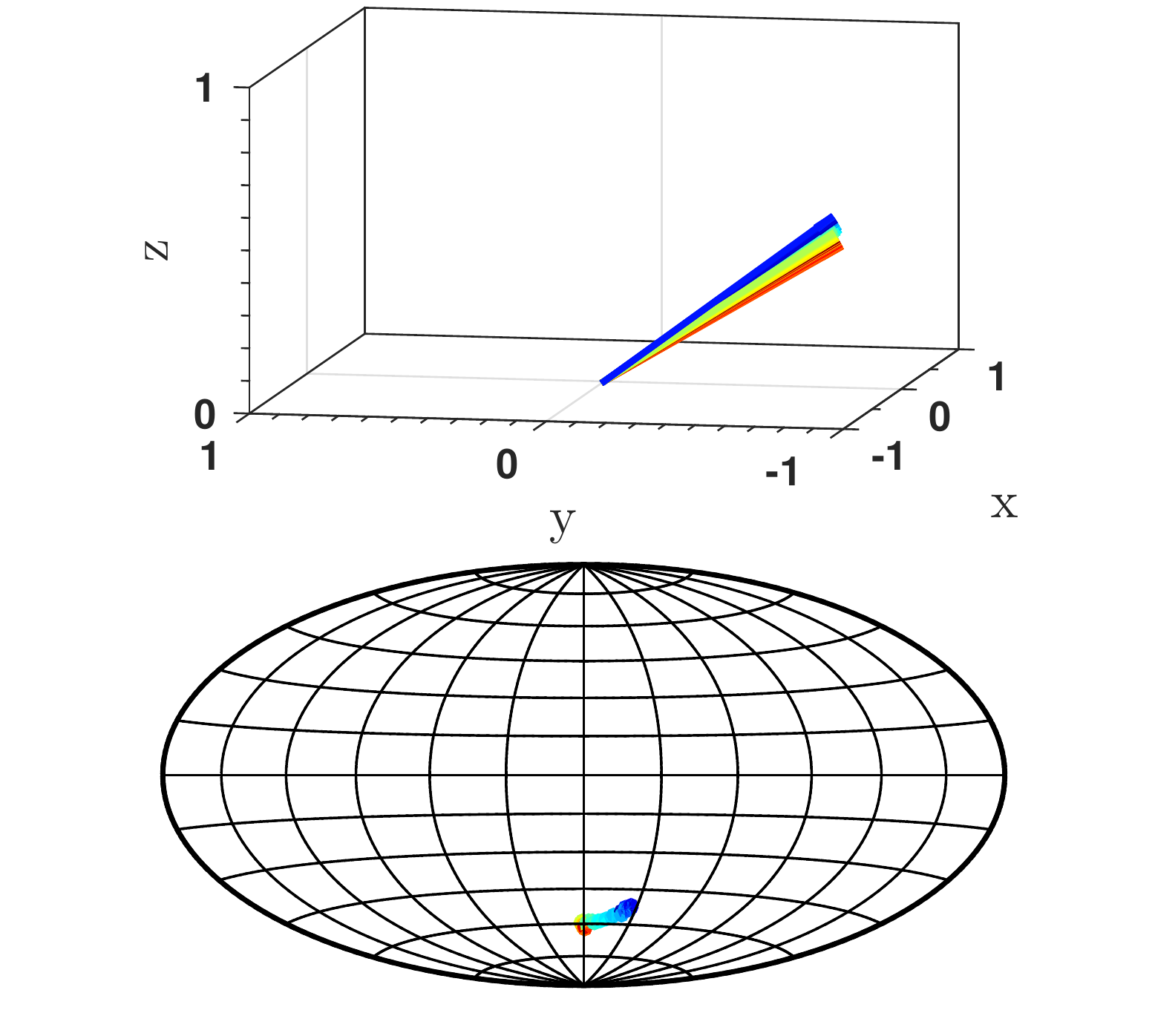}\label{Fig.4.d}}
  \subfigure[$\theta$=75$\degree$]{\includegraphics[width=0.45\textwidth]{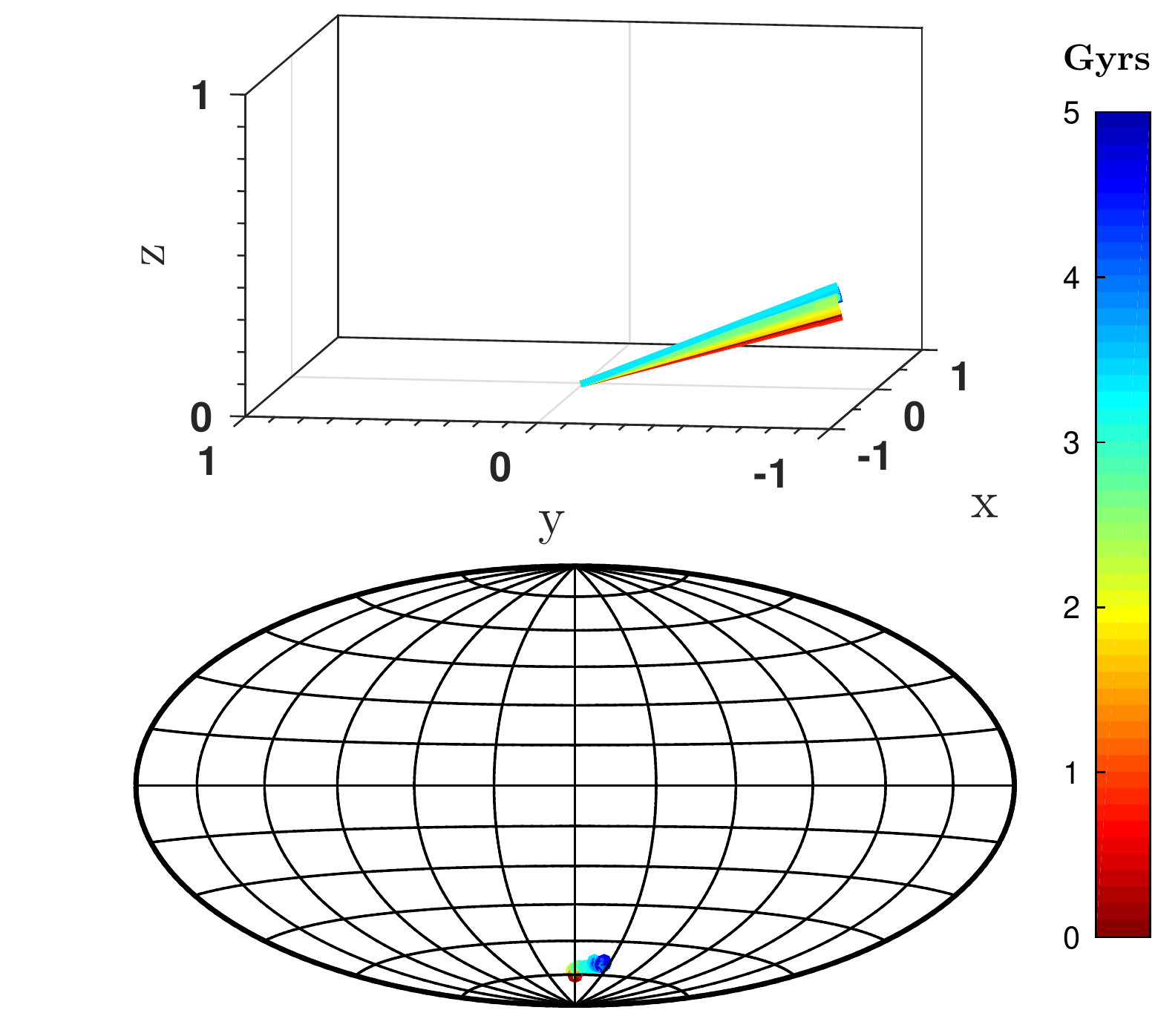}\label{Fig.4.e}}  
\caption{Variation of normals of best fit planes and their Aitoff
projection - varying inclines for $q=0.8$ of the M31 halo, for
satellites with $10^{9} \msun $ mass and 0$ \kms$ perpendicular
velocity}
\label{Fig.4}
\end{figure*}

\begin{figure*}% {normals of fitted planes} % Put Colour Bar !!!!
\centering
  \subfigure[$\theta$=15$\degree$]{\includegraphics[width=0.45\textwidth]{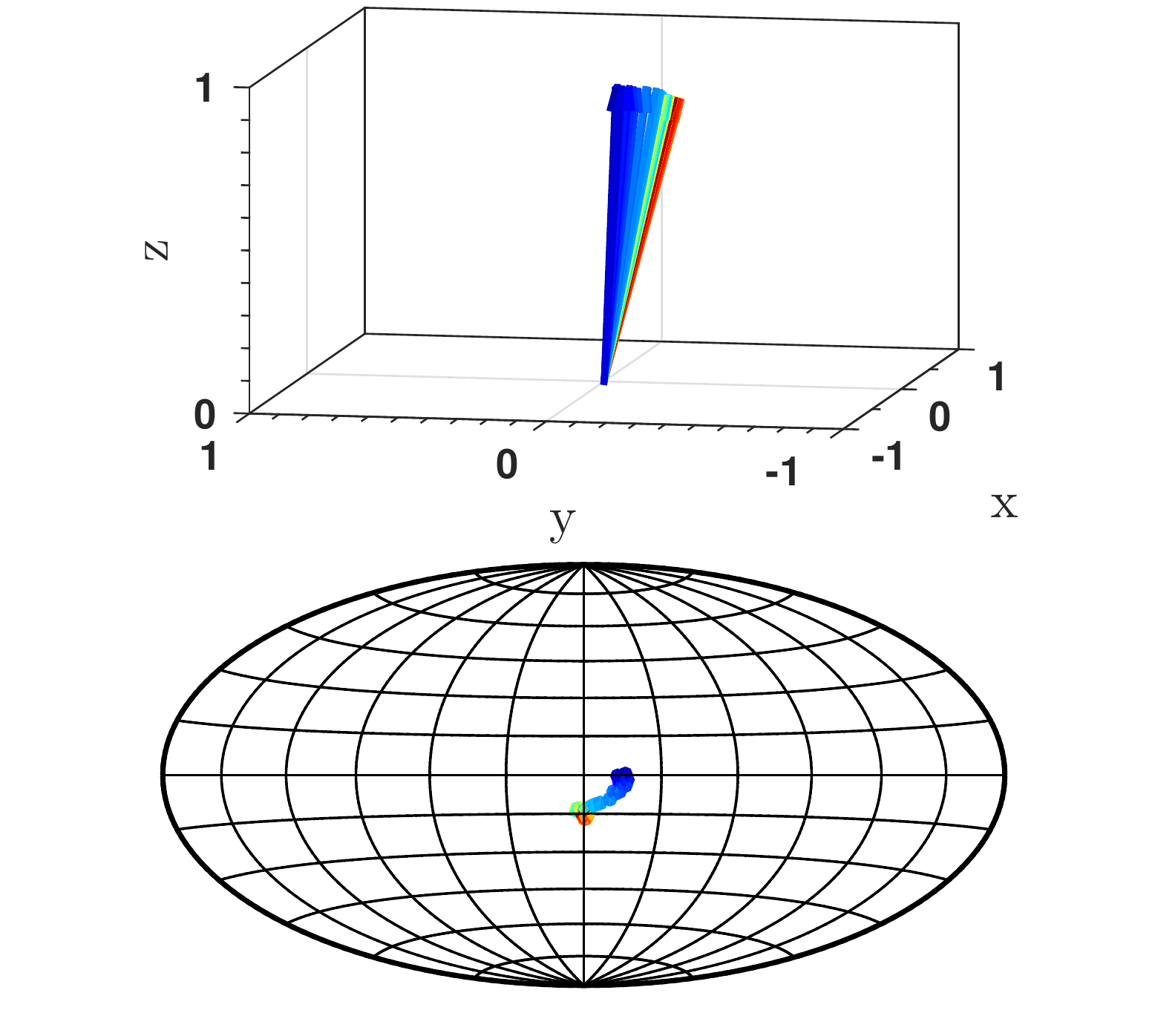}\label{Fig.5.a}}
  \subfigure[$\theta$=30$\degree$]{\includegraphics[width=0.45\textwidth]{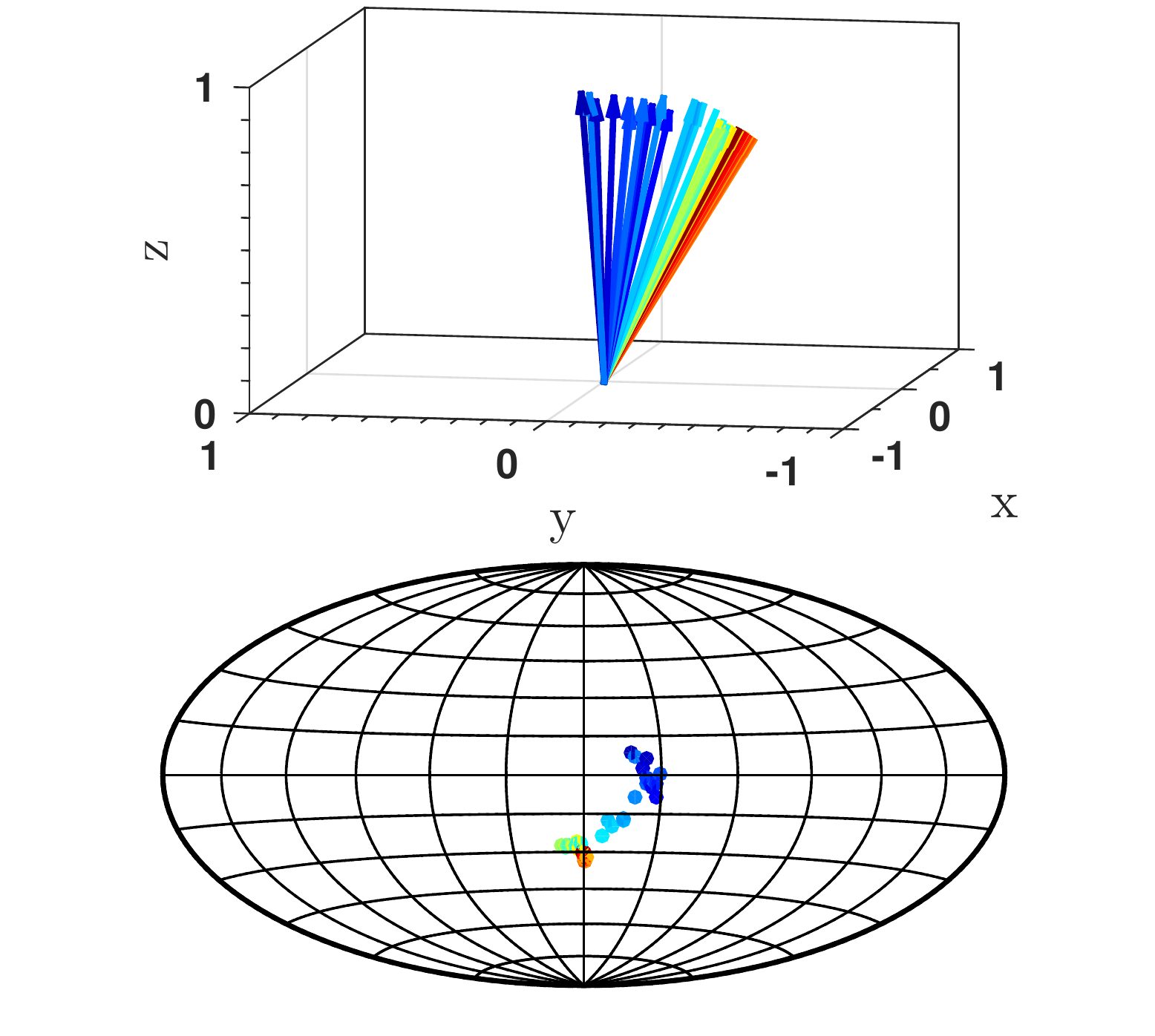}\label{Fig.5.b}}
  \subfigure[$\theta$=45$\degree$]{\includegraphics[width=0.45\textwidth]{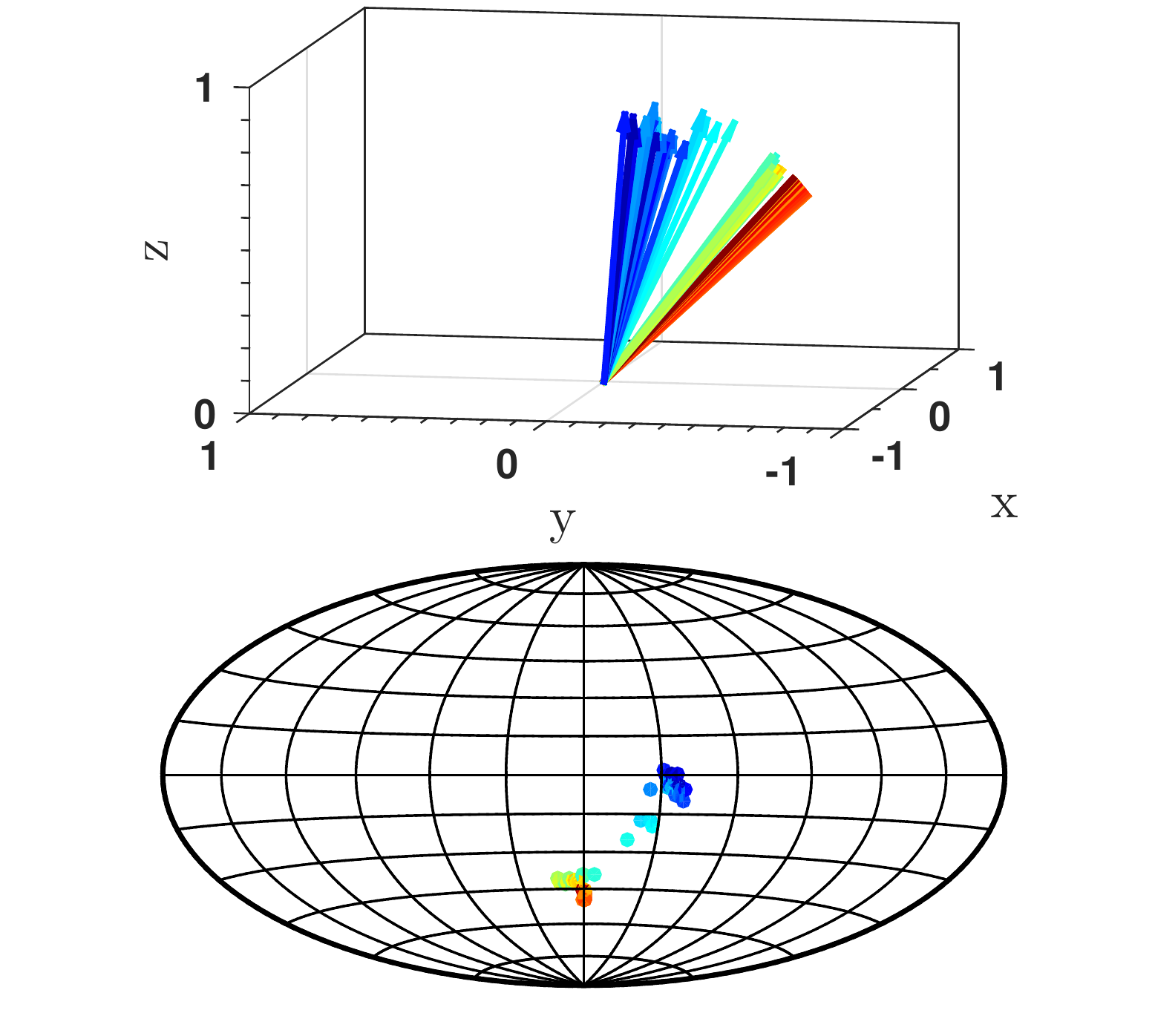}\label{Fig.5.c}}
  \subfigure[$\theta$=60$\degree$]{\includegraphics[width=0.45\textwidth]{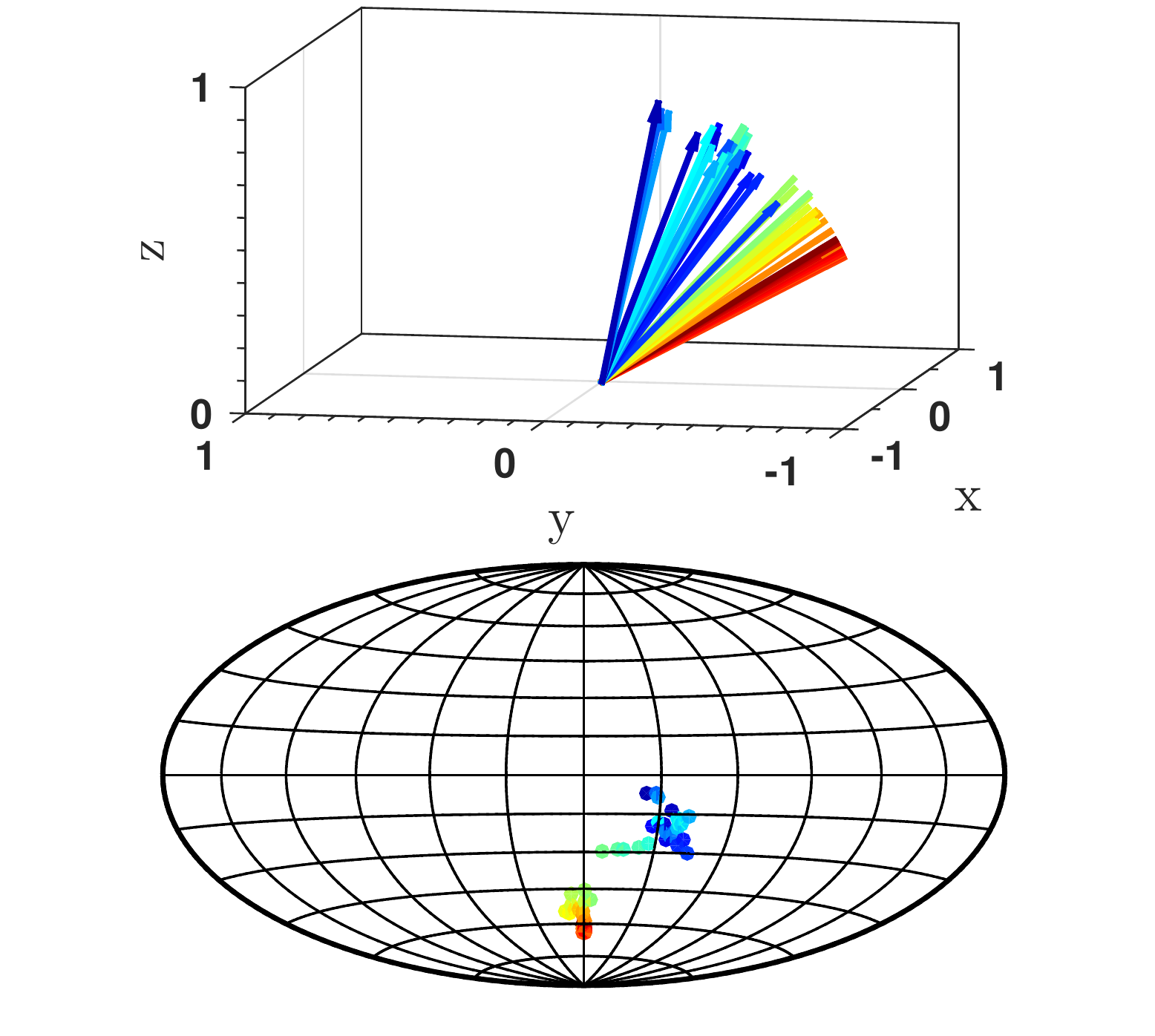}\label{Fig.5.d}}
  \subfigure[$\theta$=75$\degree$]{\includegraphics[width=0.45\textwidth]{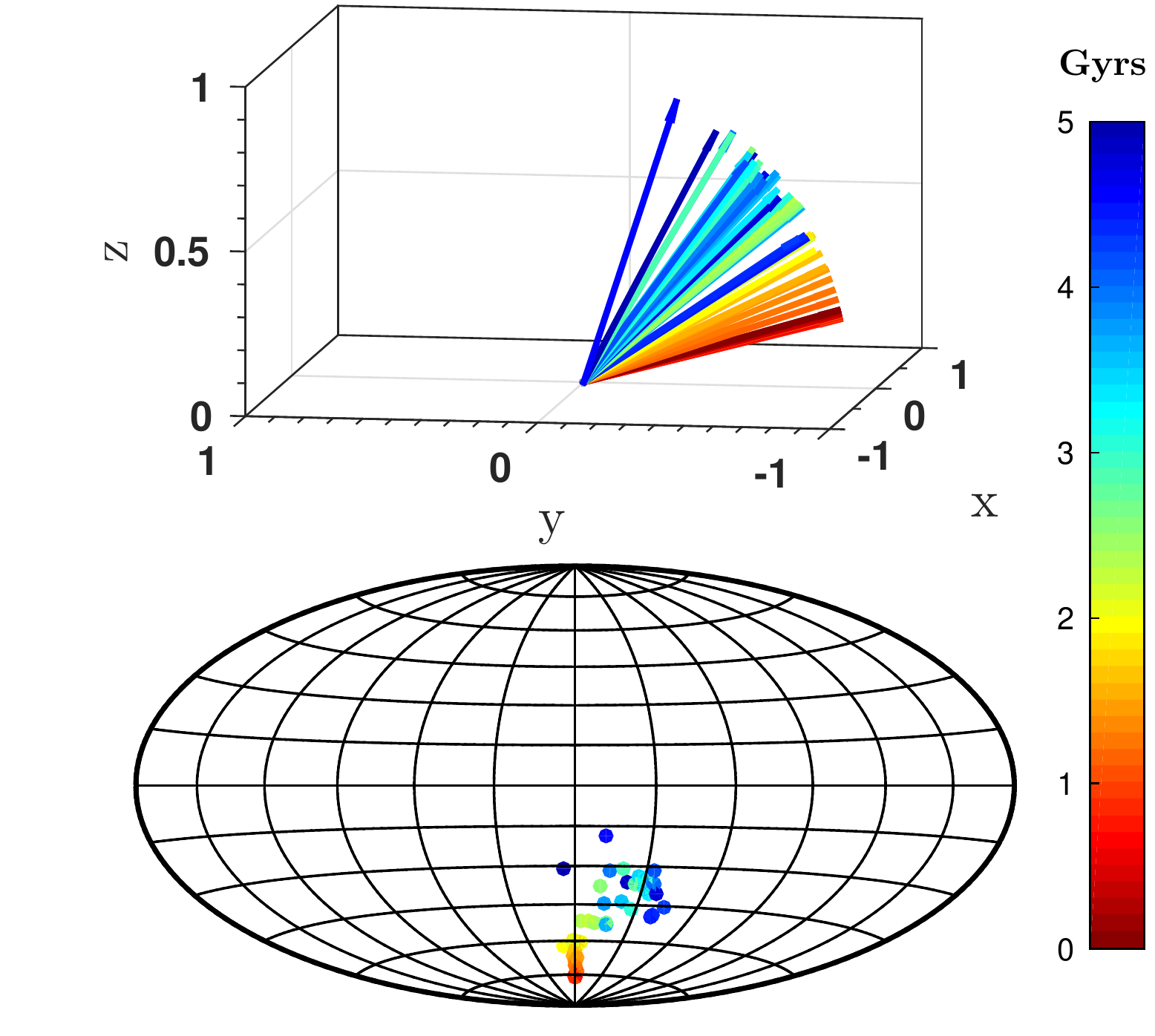}\label{Fig.5.e}}  
\caption{Variation of normals of best fit planes and their Aitoff
projection- varying inclines for $q=0.6$ of the M31 halo, for
satellites with $10^{9} \msun $ mass and 0$ \kms$ perpendicular
velocity}\label{Fig.5}
\end{figure*}

Normal vectors of best-fit planes provide a more descriptive view of
the orientation and time evolution of the best-fit planes.
Figures ~\ref{Fig.4} and ~\ref{Fig.5} show the normal direction average
of the best-fit plane for each snapshot in time for $q=0.8$ and
$q=0.6$ in Aitoff projections. A colormap for normal vectors begins at
0 Gyrs in red and proceeds on to blue as time goes to 5 Gyrs.
Fig.~\ref{Fig.4.a}-~\ref{Fig.4.e} show little or no deviation
of the normal vector from the initial position. This shows that a
plane created at any inclination will only show slight deviation from its
initial position through 5 Gyrs in a spherical halo structure.  The
next set of figures representing more flatter halo ($q=0.6$) in
Figs.~\ref{Fig.5.a}-~\ref{Fig.5.e} show that the normal vectors of the
best fit planes have a much bigger spread in their directions,
specially in the last 3 Gyrs. The placements of normal vectors in the
first 1-2 Gyrs show a unidirectional time evolution that can visually
seem like a `precessing plane', but this can be attributed to the
increasing thickness of the initial plane. Also, as best-fit models
will choose any satellite within the fitted plane range, it is
unlikely that we are seeing the precession of a single plane. For
every $q \leq 0.8$ halo the largest spread of normals over the Aitoff
projection occurs at 45$\degree$. The planes are oriented in a more
random manner as the flatness of the halo increases. Therefore, we
cannot draw a conclusion on the precession of satellite planes in
prolate/oblate halos.   
 
\section{Discussion}\label{Discn}
The above tests give us a view of how the stability of a satellite
galaxy system in an M31-like potential is affected by a few
fundamental factors. Apart from looking at the probability of seeing a
satellite galaxy on a plane at a given time, we have calculated the
average time a satellite galaxy spends as a part of a planar formation
(Tables~\ref{Table.2},~\ref{Table.3}; using the same definition as previous tests).

\subsection{Properties of Satellite Planes and Their Influences on Stability}

As far as the properties of the satellite galaxies are concerned,
velocities perpendicular to the plane exerts the most significant
influence on the stability and longevity of the plane. Even our toy
model of 30 satellite galaxies requires a perpendicular velocity
spread $\sigma \leq$ 30 $\kms$ (with $\mu = 0 \kms$)  
to produce a plane with at least 10 satellites, that survives a 5 Gyr orbital integration run. Without such a restricted velocity spread,
planes tend to have a lifetime smaller than 2 Gyrs.  
Exploration in \citet{Buck2015} of the kinematics of planes in `zoomed
in' cosmological simulations states that around 25\%\ of satellite
galaxies had very high velocity components perpendicular to the planar
formation at the time of accretion. By results presented in this paper (particularly in Fig.~\ref{Fig.1}), such high perpendicular velocities of the plane satellites are likely to produce short lived planes.
 
All of our satellites are given the same mass. Varying this value shows (Fig.~\ref{Fig.2}) that only the
self-interaction of equal-mass satellites (in a spherical halo) is not enough
to completely disperse a plane of satellites. For this to be apparent, all satellites need to be $\sim$$10^{10}$ $\msun$. But M31 satellites are spread over a wide spectrum in mass and there are only a few galaxies that have a calculated total mass close to the $10^{10} \msun$. These larger members may play a significant role in disrupting the plane, while smaller satellites ($10^{6}, 10^{7}, 10^{8}$ $\msun$) are likely to stay on an initially set plane for 5 Gyrs. To consider this variation, we ran the same tests with a spectrum of masses for the satellites (from $10^{6}$ - $10^{10}$ $\msun$). We observe that inclusion of satellite masses $\leq$ $5$x$10^{9}$ $\msun$ cause interactions that disrupt the orbits of the smaller or similar mass satellites. This results in planes with increased thickness and gives satellites a shorter average time on the planes. However, a plane of lower mass satellites (mass $\leq 10^{9} \msun$) will disperse at a lower perpendicular velocity limit than satellites with larger masses. Therefore, while plane perpendicular velocities
influence the longevity of the plane, the satellite mass has an
effect in altering the maximum perpendicular velocities required for
stable formations.  
 
Additionally, our results suggest that even satellite planes
containing galaxies of total mass $10^{9}$ $\msun$ with a $\sigma = 0\kms$,
need to be aligned with the host dark halo principal axes or to reside
in halos with a flatness $\leq$ 0.9 to remain stable through 5 Gyrs.
For halos that show an extreme change in its spherical nature ($q$ = 0.8 and 1.67 for prolate and oblate), satellites in planes that are inclined at $45\degree$ to the $x$ axis spend as
little as 1 Gyr within the $D_{\rm{rms}}$ = 15 \kpc\ range of the
initially set plane. When considering a purely observational
arrangement by least-square fitting, where the satellites are seen in
a plane by chance, an average satellite can be seen spending up to 2.5
Gyrs of the total 5 Gyrs on the best-fit plane at each snapshot. This
applies even for the most prolate/oblate halos considered. Our results agree
with \citet{Bowden2013} and their work on the quadrupole moment of the M31
dark matter halos. Their test suggest the torque on their simulated plane of satellite galaxies result in
a thicker distribution at the end of their 7 Gyr simulation. The
final `disc height' of satellites increased up to 25 \kpc\ when the
halo axes are misaligned with the stellar disc of M31 and the `disc of
satellites'.
 
To place our simulations in context
with observations, we changed our definition of a `satellite plane'-
increasing the threshold number from 10 to 15 out of 30. The trends
that appear in the set of Figure set~\ref{Fig.3} are not changed, but exaggerated.
The most notable change is that most initial formations (excluding
those on $\theta=0\degree, 90\degree$ planes) reach very low
probability of existing as planes of 15 or more satellites
($P\sim0.3$), 0.5 -1.0 Gyrs earlier than planes of 10 or more
satellites. This gives a shorter lifetime for M31-like planes.
Additionally, if we modify the plane-fitting function to fit the
thickness of the M31 plane restrictions (a 14 $\kpc$ thickness instead
of 30 $\kpc$), the longevity of the non-aligned planes in non-spherical
halos will decline further by $\sim$0.5 Gyrs for steeper inclines.
Because keeping a larger (than 1/3) fraction of the entire satellite
population for a 5 Gyrs period are more probable in spherical halos or
axis-aligned planes, these results further support the possibility of
a plane in M31 being closely aligned with a M31 dark halo axis.

Formation theories using Tidal Dwarf Galaxies \citep{1992Natur.360..715B} as building blocks for these structures, a scenario explored by many \citep[e.g.][]{Hammer2013, 2013sf2a.conf..227F, 2014MNRAS.442.2419Y}, also claim that older, higher concentration halos are more likely to host satellite planes - where the structures themselves are formed quite
early ($\geq$ 5 Gyrs ago). This allows for giant tidal streams to
coalesce into the satellites that are seen today. Considering our
results, given the non-alignment of the M31 stellar disc to the
satellite plane and a predicted triaxial halo for M31, a plane surviving longer than 5 Gyrs will require its satellite galaxies to maintain perpendicular velocities extremely restricted to the plane of galaxies. The question to ask
is if the dynamic environments of tidal tails and accretions are
capable of producing such restricted velocities. It is also difficult to make claims on tidal galaxies as origins of these planes from our numerical model. A shared origin would influence intrinsic properties such as velocity in tidal satellite galaxies, whereas our model's positions are chosen randomly and circular velocities are assigned to keep galaxies on their initially set planes. Our model also refers to planes that are already formed, and therefore cannot sufficiently speculate on formation of planes. \citet{Bowden2013}
also consider misaligned planes as good tracers of the underlying
dark matter structure. This, taken into consideration with the fact that the M31 plane forms $\sim 50\degree$\ angle with the stellar disc of M31 create another set of questions on the dark halo-stellar disk  alignment of M31 and stability of the M31's VTPoS. %This, taken into consideration with the facts that (a) the M31 plane forms an $81\degree$\ angle with the stellar disc of M31 and (b) \citet{Vera-Ciro2011} gives a higher likelihood of MW-like stellar halos aligning with a dark halo axis, create another set of questions on the stability and life-time of the M31 VTPoS. \color{red}???\color{black}
 
\section{Conclusions}\label{Concln}
Explanations for the M31 plane of satellite galaxies have taken
various avenues. By considering a numerical model of an M31 and MW
potential, we conducted orbital integration on 30 satellite galaxies
with varying environments and properties. We set satellite galaxies on
an initial plane to calculate the probabilities for stability and
longevity in a simulation time of 5 Gyrs. The following properties
were varied both individually and in combination: satellite velocities
perpendicular to the plane, mass of the satellites, inclination of
satellite plane to the triaxial host halo, and the non-spherical
nature of the host halo. From our findings so far, the following
conclusions can be inferred by the statistics obtained from each set
of simulations.\\
 
[1] A long-lived plane of satellite galaxies must be dynamically cold
- with the magnitude of the velocity vector perpendicular to the plane
being smaller than $20 \kms$. A satellite plane with a perpendicular
velocity distribution with $\sigma$ = 50$\kms$ disperses to contain
about half the initial count of satellite galaxies in around 2 Gyrs.
\\
    
[2] The shape of the host halos and the inclination of an initial
plane of satellites, both impart great influence on
the stability of the plane. Unless a host halo is nearly
spherical, it is highly unlikely that a plane that is off a major or
minor axis by an angle larger than 30$\degree$ has existed for a
period longer than 1 Gyr. \\

[3] There are no signs of precession of one single plane continuously
over 1 Gyr. Best-fit planes show that expanding planes might show
precession-like movement, but no long term order is found in even the
simplest scenarios tested. \\
 
These deductions can provide us with insights into an M31-like system
of satellites. It is unlikely that the plane of
galaxies currently seen is very old ($\leq$ 4 Gyrs) unless the
system has velocities restricted to the plane. If the proper motions
of the satellites show a system with a large variation of velocities,
we can conclude that the plane is a rather young formation that is
dispersing or will do so in a short (galactic) time span. The unusual
thinness of the M31 plane requires either an extraordinarily well-aligned initial plane of galaxies to be older than a few Gyrs, or
dictates that this is a rather recent formation.

The surrounding environment of dark matter subhalos and interactions with them and perturbations by larger satellite galaxies of M31 (e.g. M33) will be addressed subsequently.
 
\section*{Acknowledgments}
N. F. acknowledges the Dean's International Postgraduate Scholarship of the Faculty of Science, University of Sydney.  
 
\bibliographystyle{mn2e}
\bibliography{Master}

\label{lastpage}
 
\end{document}